\documentclass{aa}  
\usepackage{graphicx}
\usepackage[]{natbib}
\bibpunct{(}{)}{;}{a}{}{,} 
\usepackage[varg]{txfonts}
\usepackage{multirow}
\usepackage{booktabs}
\usepackage{mathtools}  
\usepackage{txfonts}
\usepackage{bigstrut}
\usepackage{hyperref}

\newcommand*\diff{\mathop{}\!\mathrm{d}}

\begin{document}

   \title{Gamma rays from jets interacting with BLR clouds in blazars} 

   \titlerunning{Cloud-jet interactions}
   \authorrunning{S. del~Palacio, V. Bosch-Ramon, G. E. Romero}

   \author{
   S. del~Palacio \inst{1,2} 
   \and V. Bosch-Ramon \inst{3}
   \and G. E. Romero \inst{1,2}
    }

   \institute{
  Instituto Argentino de Radioastronom\'{\i}a (CCT-La Plata, CONICET; CICPBA), C.C.5, 1894, Villa Elisa, Argentina.  
  \and Facultad de Ciencias Astron\'omicas y Geof\'{\i}sicas, Universidad Nacional de La Plata, Paseo del Bosque, B1900FWA La Plata, Argentina.
  \and Departament de F\'isica Qu\`antica i Astrof\'isica, Institut de Ci\`encies del Cosmos (ICCUB), Universitat de Barcelona, 
  IEEC-UB, Mart\'i i Franqu\`es 1, E08028 Barcelona, Spain }

   \date{Received - ; accepted - }

  \abstract
   {The innermost parts of powerful jets in active galactic nuclei are surrounded by dense, high-velocity clouds from the broad-line region, which may penetrate into the jet and lead to the formation of a strong shock. Such jet-cloud interactions are expected to have measurable effects on the $\gamma$-ray emission from blazars.}
   {We characterise the dynamics of a typical cloud-jet interaction scenario, and the evolution of its radiative output in the 0.1-30~GeV energy range, to assess to what extent these interactions can contribute to the $\gamma$-ray emission in blazars.}
   {We use semi-analytical descriptions of the jet-cloud dynamics, taking into account the expansion of the cloud inside the jet and its acceleration. Assuming that electrons are accelerated in the interaction and  making use of the hydrodynamical information, we then compute the high-energy radiation from the cloud, including the absorption of $\gamma$-rays in the ambient photon field through pair creation.}
   {Jet-cloud interactions can lead to significant $\gamma$-ray fluxes in blazars with a broad-line region, in particular when the cloud expansion and acceleration inside the jet are taken into account. This is caused by 1) the increased shocked area in the jet, which leads to an increase in the energy budget for the non-thermal emission; 2) a more efficient inverse Compton  cooling with the boosted photon field of the broad-line region; and 3) an increased observer luminosity due to Doppler boosting effects.}
   {For typical broad-line region parameters, either (i) jet-cloud interactions contribute significantly to the persistent $\gamma$-ray emission from blazars or (ii) the broad-line region is far from spherical or the fraction of energy deposited in non-thermal electrons is small.}
 
   
   \maketitle
%
%


\section{Introduction}\label{sec:intro}


The majority of the $\gamma$-ray sources detected by the \textit{Fermi}/LAT instrument are associated with the so-called blazars, a type of active galactic nuclei (AGNs) with powerful jets pointing almost directly at us \citep{Fermi2015_3FGC, Fermi2015_AGNcat}. The blazar population is made up by two subclasses of objects: the flat-spectrum radio quasars (FSRQs) and the Bl Lac objects. The former present a prominent environment surrounding the jet, including dense,  high-velocity clouds located at the broad-line region (BLR). This work focuses on a mechanism that may contribute to the observed high-energy $\gamma$-ray (HE) emission from FSRQs.

The spectral energy distribution (SED) of a blazar typically consists of two broad `humps', one at low energies (that extend to X-rays at most), presumably synchrotron in  nature, and one in $\gamma$ rays, most likely produced by inverse Compton (IC) radiation\footnote{The SED can show additional components of thermal emission from the disc, the dusty torus, and the BLR.}. A detailed study of the emission properties of the blazar population detected by {\it Fermi} has been recently published by \cite{Ghisellini2017}. Based on their analysis of the SED of 448 {\it Fermi} FSRQs, they report the following behavior in $\gamma$ rays: i) the Compton dominance (i.e. the ratio between the $\gamma$-ray and the synchrotron emission) increases with the blazar luminosity from 0.5 for $L_\gamma \sim 10^{44}$ to 15 for $L_\gamma \sim 10^{48}$; ii) the $\gamma$-ray slope is almost constant;
and iii) the Compton peak frequency varies slightly from $5 \times 10^{21}$ to $9 \times 10^{20}$~Hz (i.e. from to $\sim 21$ to $4$~MeV) for an increase in luminosity of 4 orders of magnitude.
Here we investigate the $\gamma$-ray emission, focusing in particular on the HE range, for the scenario of a jet-BLR cloud interaction (JCI) in a blazar \citep[e.g.][]{Barkov2012_Jet-Star,Khangulyan2013,Aharonian17}. \cite{Araudo2010} showed that BLR clouds can penetrate inside the AGN relativistic jet leading to the formation of strong shocks, in which non-thermal (NT) particles can accelerate through diffusive shock acceleration (DSA) and generate $\gamma$-ray emission \citep[for similar scenarios, also see  e.g.][]{dar97,beall99,beall02}. We consistently model the dynamics of this interaction, the evolution of the NT particle population in the emitting cloud (i.e. the shocked jet material surrounding the shocked cloud), and its expected $\gamma$-ray output, including Doppler boosting effects. The $\gamma$-ray emission from external Compton (EC) interactions of relativistic electrons in the jet with BLR photons has been studied by several authors \citep[e.g.][]{Tavecchio2008, Finke2016}. 
In such works neither the nature of the dynamical mechanism leading to the acceleration of NT particles nor its evolution is specified. Thus, here we aim to explore the role of the JCI scenario  as a potential contributor to the overall blazar HE emission. Our findings can be put in contrast with the recent results by \cite{costamante18} who suggest that the role of the BLR in blazar $\gamma$-ray emission is rather minor.


\section{Physical scenario}\label{sec:scenario}

\begin{figure}
    \centering
    \includegraphics[width = 0.67 \linewidth]{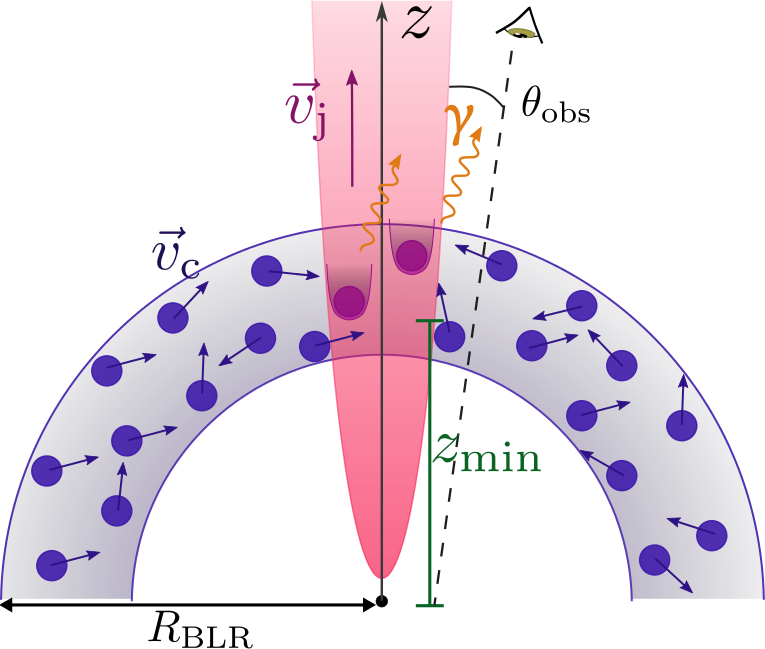} 
    \caption[]{Schematic representation of the BLR, the JCIs, and some of the relevant quantities defined in \S\ref{sec:dynamics}.}
    \label{fig:sketch}
   \end{figure}

The BLR has a large number of relatively dense clouds that can possibly penetrate into the jet. As shown by \cite{Araudo2010}, the cloud volume takes only a short timescale to fully enter the jet, and the impact of the relativistic jet on the cloud surface leads to the formation of two shocks. One shock propagates in the cloud, whereas the other (the bow-shock) propagates in the jet material; the latter is suitable for the acceleration of NT particles that can produce high-energy radiation\footnote{In the relativistic regime (when the cloud has been already accelerated by the jet) the energy transfer in both shocked regions (jet and cloud) is similar.}. A sketch is presented in Fig.~\ref{fig:sketch}.

\subsection{Broad-line region} \label{subsec:BLR}

Neither the formation process of the BLR nor its geometry are clearly understood. Several works favour a spherical geometry for the BLR \citep{Kaspi1999, Kaspi2005, Kaspi2007, Liu2006}, 
whereas others consider a bowl-like geometry \citep{Gaskell2009, Goad2012}. \cite{Grier2013} presented observations that are compatible with inflows and outflows from either an inclined disc or a spherical shell geometry. In general, both the number density of clouds ($N_\mathrm{c}$) and their size ($R_\mathrm{c,0}$) are considered to depend on the radial distance as $N_\mathrm{c} \propto r^{-p}$, with $p=1.5-2$, and $R_\mathrm{c,0}\propto r^q$, $q=1/3-1$, depending on the geometry. Typical values of the size and density of the BLR clouds are $R_\mathrm{c,0} \sim 10^{13}$~cm and $n_\mathrm{c,0} \sim 10^{10-11}$~cm$^{-3}$ \cite[e.g.][and references therein]{Netzer2015}. 

For simplicity, in this work we assume that the BLR is spherical and with an outer radius $R_\mathrm{BLR}$ of which the clouds fill only a small fraction of the volume. 
We further assume that all BLR clouds have a similar size and density, fixing $R_\mathrm{c,0}=10^{13}$~cm and $n_\mathrm{c,0}=10^{10}$~cm$^{-3}$. From these considerations, the total number of BLR clouds is $N_\mathrm{c,tot} \gtrsim 10^6$ for $R_\mathrm{BLR} \gtrsim 10^{17}$~cm \citep[see also][]{Dietrich1999}, and the cloud mass (assuming hydrogen), $M_\mathrm{c}= 7\times 10^{25}$~g. A further analysis regarding the implications of the size and spatial distribution of the BLR clouds is discussed in \S\ref{sec:interactions}. The BLR radiation field can be regarded as isotropic within $R_\mathrm{BLR}$ given that the number of clouds is very large, and assuming that they are distributed uniformly inside the BLR \citep[e.g.][]{Donea2003}.

In our simplified scenario, the properties of the BLR are determined by its outer radius $R_\mathrm{BLR}$ and its luminosity $L_\mathrm{BLR}$. Most of the emission produced by the BLR is a reflection of the disc photons that ionise and excite the atoms in the BLR. It is then sensible to assume that the BLR luminosity, $L_\mathrm{BLR}$, is a fraction of the disc luminosity, $L_\mathrm{d}$. Moreover, the jet power also scales with the  disc luminosity as $L_\mathrm{d} \propto L_\mathrm{j}^a$, where $a \approx 1$ \citep[e.g.][]{Ghisellini2014} and $L_\mathrm{BLR} \approx 0.5-0.1 L_\mathrm{j}$ \citep{Ghisellini2009}. The size of the BLR is related to the BLR luminosity as $R_\mathrm{BLR} \propto L_\mathrm{BLR}^{0.5} \propto L_\mathrm{j}^{0.5}$. 
We note that the adopted value of $a$ affects the feasibility of JCIs: values of $\alpha > 1$ \citep[e.g.][]{zhang15} make it more likely that BLR clouds penetrate powerful jets since JCIs require that $R_\mathrm{BLR} > z_\mathrm{min}$ (see Eq.~\ref{eq:z_min}  and  Sect.~\ref{sec:dynamics}). 


\section{Cloud dynamics and non-thermal processes}
   
In this section we present the formulas used in the calculation of the dynamical quantities, and our radiative code. We focus here on leptonic processes, namely synchrotron and IC, as in this particular scenario the target densities are too low for relativistic Bremsstrahlung, proton-proton emission, and photo-meson production to be efficient, and it is difficult to accelerate protons up to energies high enough for proton synchrotron to be relevant \citep[see e.g.][for similar scenarios with hadronic emission]{Barkov2012_Jet-Cloud,Khangulyan2013}. Throughout this work, un-primed quantities  refer to the laboratory frame (LF), primed quantities  refer to the cloud (co-moving) frame (CF), and quantities with a hat  refer to the observer frame (OF). The shocked fluid is assumed to move with the accelerating cloud, and therefore the shock frame is at rest with respect to the CF. We define $\beta = v/c$. 
We focus on the advanced stages of the JCI, which are particularly interesting for the case of blazars. The main effects to take into account are the following: i) the area of the bow-shock increases because of the expansion of the cloud inside the jet; ii) the IC cooling becomes more efficient when the cloud reaches relativistic velocities because the BLR photon field is  Doppler-boosted in the CF; and iii) the emitted radiation is greatly enhanced in the OF by Doppler boosting. All these phenomena lead to a substantially larger $\gamma$-ray emission than that estimated by \cite{Araudo2010}, in which relativistic effects were not considered, and the clouds were unperturbed. 

\begin{table}
\caption{Parameters for a typical JCI.} 
\centering
 \begin{tabular}{lcc}
  \toprule
  \textbf{Parameter}                &   \textbf{Model (canonical jet)}              \\
  \midrule
  Jet luminosity [erg s$^{-1}$]     &   $L_\mathrm{j} = 2.5\times 10^{46} \,^{(a)}$       \\
  Jet Lorentz factor                &   $\Gamma_\mathrm{j} = 13 \, ^{(b)}$                      \\
  Black hole mass [M$_\odot$]           &       $M_\mathrm{BH} = 6\times 10^{8} \, ^{(b)}$       \\
  Disc luminosity [erg s$^{-1}$]        &       $L_\mathrm{d} = 0.1 L_\mathrm{j}$ $^{(a)}$        \\
  BLR luminosity [erg s$^{-1}$]]        &       $L_\mathrm{BLR}= 0.1 L_\mathrm{d}$ $^{(c)}$        \\
  BLR radius [cm]                               &       $R_\mathrm{BLR}= 1.6\times 10^{17} \, ^{(c)}$ \\
  \midrule
  BLR cloud radius [cm]                 &       $R_\mathrm{c,0}= 10^{13}$ $^{(d)}$                      \\
  BLR cloud density [cm$^{-3}$]     &   $n_\mathrm{c,0}= 10^{10}$ $^{(d)}$                        \\
  BLR cloud velocity [cm s$^{-1}$]      &       $v_\mathrm{c,0}= 10^{9}$                                     \\
  BLR cloud mass [g]                &   $M_\mathrm{c}= 7\times 10^{25}$                         \\
  \midrule
  Poynting-to-kinetic energy flux       &       $\eta_B = 1$ \\ 
  NT fraction (electrons)               &       $\xi_\mathrm{e} = 10^{-1}$                               \\    
  Acceleration efficiency                   &   $\eta_\mathrm{acc} = 0.1$                                        \\    
  Observing angle [rad]                     &   $\theta_\mathrm{obs}= \theta_\mathrm{j}$        \\    
  \bottomrule
  \end{tabular}
  \tablefoot{$^{(a)}$ \citep{Ghisellini2014}; $^{(b)}$ \citep{Ghisellini2015}; $^{(c)}$ \citep{Ghisellini2009}; $^{(d)}$ \citep{Netzer2015}.}
 \label{tab:parameters}
\end{table}


\subsection{Dynamics} \label{sec:dynamics}

We present here the main characteristics of a JCI and show that for conservative assumptions a JCI is likely to occur. The following quantities are required for the analysis: the BLR cloud velocity outside the jet (roughly its Keplerian velocity orbiting the central black hole: $v_\mathrm{c,0} \approx \sqrt{G M_\mathrm{BH}/R_\mathrm{BLR}}\approx 10^9(M_\mathrm{BH}/10^9 \mathrm{M}_\odot)^{1/2}(R_{\rm BLR}/10^{17}\,{\rm cm})^{-1/2}$~cm~s$^{-1}$), the initial cloud number density ($n_\mathrm{c,0}$) and radius ($R_\mathrm{c,0}$); the jet luminosity ($L_\mathrm{j}$), Lorentz factor ($\Gamma_\mathrm{j}$), penetration height ($z_\mathrm{j}$), radius ($R_\mathrm{j}$), density ($\rho_\mathrm{j}$), and aspect ratio  (parametrised\footnote{For simplicity, the jet region is taken as conical regardless of its actual shape.} as $\theta_\mathrm{j}=R_\mathrm{j}/z_\mathrm{j}$); the disc luminosity ($L_\mathrm{d}$), radius ($R_\mathrm{d}$), and temperature ($T_\mathrm{d}$); the torus luminosity ($L_\mathrm{t}$), inner radius ($R_\mathrm{t}$), and temperature ($T_\mathrm{t}$); and the BLR luminosity ($L_\mathrm{BLR}$), radius ($R_\mathrm{BLR}$), and characteristic temperature ($T_\mathrm{BLR}$). The typical values for these parameters are given in Table~\ref{tab:parameters}. 

For clouds to fully penetrate into the jet it is required that $v_\mathrm{sh} < v_\mathrm{c,0}$. If this condition is not fulfilled, the cloud becomes disrupted in the shear layer between the jet and the surrounding medium, and it is dragged along the downstream direction without penetrating further into the jet (before being significantly accelerated, so no strong direct interaction would take place). From this condition a minimum penetration height to produce a strong JCI can be derived \citep{Araudo2010}:
\begin{align} \label{eq:z_min}
 z_\mathrm{min} &= \frac{1}{v_\mathrm{c,0} \theta_\mathrm{j}} \sqrt{ \frac{L_\mathrm{j}}{\pi c m_\mathrm{p} n_\mathrm{c,0}} } \\ \nonumber
   &\approx 5\times 10^{16} \left(\frac{0.1}{\theta_\mathrm{j}}\right) \left(\frac{10^9}{v_\mathrm{c,0}}\right) \sqrt{ \left(\frac{10^{10} }{n_\mathrm{c,0}}\right) \left(\frac{L_\mathrm{j}}{10^{46}}\right)} \; \mathrm{cm},
\end{align}
all quantities being in cgs units. The maximum interaction height we are interested in is $z_\mathrm{max} \approx R_\mathrm{BLR} \approx 3 z_\mathrm{min}$, although the ratio $z_\mathrm{max}/z_\mathrm{min}$ depends on $L_\mathrm{j}$ for $a\neq 1$ (see $\S$\ref{subsec:BLR}). We take a location for the JCI at an average value (in terms of radiative output) of $z_\mathrm{c,0} = \sqrt{z_\mathrm{max}\,z_\mathrm{min}}$.

Once we have the initial conditions set, we can calculate the dynamical evolution of the shocked cloud inside the jet. The jet impact initially compresses the cloud in the jet direction of motion; the cloud quickly gets disrupted and begins accelerating along the jet and expanding \citep{Perucho2017,Zacharias2017}. We can calculate the shocked cloud acceleration by numerically solving a differential equation for its Lorentz factor, $\Gamma_\mathrm{c}$. The quantities required in each time-step are calculated as follows. The jet radius is $R_\mathrm{j} = z_\mathrm{c} \theta_\mathrm{j}$, and its ram pressure in the LF is $P_\mathrm{j} = L_\mathrm{j}/(c \pi R_\mathrm{j}^2)$. Equation~A3 from \cite{Barkov2012_Jet-Star} gives the jet ram pressure in the CF:
\begin{equation}
 P'_\mathrm{j} = P_\mathrm{j} \Gamma_\mathrm{c}^2 \left(\beta_\mathrm{j}-\beta_\mathrm{c}\right)\left(1-\beta_\mathrm{j} \beta_\mathrm{c}\right)\,.
 \label{eq:P_j}
\end{equation}
This equation applies to a jet dominated by Poynting flux, although for a matter dominated jet the results are very similar in the relativistic regime when emission matters the most.
The electromagnetic energy flux density in the jet is $q' = P'_\mathrm{j} c$ \citep{Barkov2012_Jet-Star}, and the injected power in the shock is 
\begin{equation} 
    L'_\mathrm{inj} = q' \pi R_\mathrm{c}^2 \, .
\end{equation}

We consider that, to the first order, the shocked cloud can be modelled as a relatively spherical cloud that expands isotropically with a velocity equal to the sound speed in the cloud\footnote{This rough approximation does not significantly affect the results because of three factors: in the relativistic regime the cloud expansion is slowed down by time dilation, the cloud differential velocity along the jet in the LF is small, and the jet lateral pressure becomes more relevant. In fact, in \cite{Barkov2012_Jet-Star} and \cite{Khangulyan2013} the cloud radius was taken to be constant.}
\begin{equation}
  R_\mathrm{c}(t') = R_\mathrm{c,0} + \int_0^{t'} c_\mathrm{s}(\tilde{t'}) \diff \tilde{t'} \,, 
  \quad c_\mathrm{s} = \sqrt{\frac{\gamma_a P'_\mathrm{j}}{\rho' h'}}\,,
 \end{equation}
where $h'$ is specific enthalpy, $\rho'$ is the shocked cloud mass density, and $\gamma_a = 4/3$ is the adiabatic coefficient of a relativistic, monoatomic ideal gas. 
The cloud mass density 
is computed as $\rho' = M_\mathrm{c}/V'_\mathrm{c}$, with $V'_\mathrm{c}\sim (4\pi/3)R_\mathrm{c}^3$, i.e. taking the cloud as spherical in the CF. The specific enthalpy is $h' = 1 + 4 P'_\mathrm{c}/(\rho' c^2)$, where $P'_{\rm c}$ is the shocked cloud pressure, taken equal to the jet ram pressure in the CF ($P'_\mathrm{j}$). The lateral expansion of the cloud leads to an increase in the area of the shocked section of the jet.

We solve the differential equation for $\Gamma_\mathrm{c}$ given by \cite{Barkov2012_Jet-Star} using a first-order Euler method with a non-uniform (logarithmic) step in time,
\begin{equation}
 \Gamma_\mathrm{c}(t) = \Gamma_\mathrm{c}(0) + \int_0^t\frac{\pi R_\mathrm{c}^2 v_\mathrm{c} q'}{M_\mathrm{c} c^3}(1-f_\mathrm{NT}) \diff \tilde{t}\,,
\end{equation}
where the factor $(1-f_\mathrm{NT})$ was incorporated to take into account the back-reaction due to the emission of photons: if a fraction $f_\mathrm{NT} \sim 0.1$ of the injected jet power goes into relativistic particles, and most of this energy (and therefore momentum) is emitted in the direction of motion (as  is the case under the JCI conditions considered), a radiation pressure $\sim f_\mathrm{NT}$ $\times$ the total jet pressure is then exerted opposite to the cloud acceleration. The time intervals in the CF and the LF relate via $\diff t' = \diff t/\Gamma_\mathrm{c}$, whereas for the OF we have \citep[e.g.][]{Rybicki1979} $\diff{\hat{t}} = \diff t (1-\beta_\mathrm{c} \cos{\theta_\mathrm{obs}}) = \diff t / (\Gamma_\mathrm{c} \delta_\mathrm{c})$, where $\theta_\mathrm{obs}$ is the angle between the line of sight and the jet axis (considered to be equal to $\Gamma_\mathrm{j}^{-1}$; see Fig.~\ref{fig:sketch}) and $\delta_\mathrm{c}~=~\left[(1-\beta_\mathrm{c} \cos{\theta_\mathrm{obs}})\Gamma_\mathrm{c}\right]^{-1}$ is the Doppler boosting factor. When $\Gamma_\mathrm{c} \rightarrow \Gamma_\mathrm{j}$, we get $\delta_\mathrm{c} \rightarrow \Gamma_\mathrm{j}$.

Following \cite{Barkov2012_Jet-Star}, we define the parameter $D$ as the dimensionless inverse mass of the cloud:
\begin{equation} 
    D = \frac{P_\mathrm{j} \pi R_\mathrm{c}^2 z_\mathrm{c}}{4 c^2 M_\mathrm{c} \Gamma_\mathrm{j}^3}\,.
\end{equation}
Previous works, such as \cite{Barkov2012_Jet-Star}, consider a relativistic motion of the obstacle, so $D$ has to be derived for mildly relativistic velocities in order to be comparable with the results obtained by those authors. This means that in order to make a valid comparison, we have to evaluate $R_{\rm c}$ when the cloud has expanded and accelerated significantly. 
This leads to $D \gtrsim 1$ for a typical JCI, which implies a rather quick dynamical evolution of the cloud inside the jet \citep{Khangulyan2013}: the distance that the cloud travels along the jet during its evolution is smaller than $R_\mathrm{BLR}$. Thus, the external radiation fields typically do not change substantially with respect to those at the penetration distance. 
The cloud acceleration time (in the LF) is $t_\mathrm{acc,c} \sim z_\mathrm{c,0}/(c\,D) \sim 10^6$~s for $z_\mathrm{c,0}\sim 10^{17}$. The NT activity of the cloud is expected to have a narrow peak in the OF: taking the cloud Lorentz factor $\Gamma_\mathrm{c} \lesssim \Gamma_\mathrm{j}$, then $\hat{t}_0 \sim t_\mathrm{acc,c}/(\delta \Gamma_\mathrm{c}) \sim 1.5 \times 10^4$~s.

Finally, we estimate the magnetic field in the CF as \cite{Ghisellini2009} by assuming that the jet magnetic pressure is a fraction $\eta_B$ of the jet ram pressure:
\begin{equation}\label{eq:B}
 \frac{{B'}^2}{8\pi} = \eta_B P'_\mathrm{j}\,.
\end{equation}
Throughout this work we adopt $\eta_B=1$ as jets are expected to be rather magnetised close to their base \citep{Khangulyan2013}, but our results are not very sensitive to $\eta_B$. 


\subsection{Radiation fields}\label{subsec:rad_fields}

There are multiple radiation fields present in the inner region of blazars, such as those produced by the BLR, the torus, and the disc. Nonetheless, for an emitting relativistic cloud, moving in the direction of the jet, and located at a height $z \sim R_\mathrm{BLR}$, the only relevant radiation field for IC interactions in the JCI is the one from the BLR. This can be seen by comparing the energy density of the different radiation fields \mbox{\cite[e.g. Fig.~2 in][]{Ghisellini2009}}. In the CF the disc photon field is de-boosted, whereas the BLR photon field is boosted; the torus photon field is also boosted, but the energy density of the torus photon field is $\sim 100$ times smaller than the BLR value at $z \sim R_\mathrm{BLR}$. 
The energy density of the BLR photon field in the CF is related to the energy density in the LF via $U'_\mathrm{BLR}\sim \delta_\mathrm{BLR}^2 U_\mathrm{BLR}$, where $\delta_\mathrm{BLR}$ is the Doppler factor \cite[e.g.][]{Dermer2009}. The value of $U_\mathrm{BLR}$ can be estimated as
\begin{equation} \label{eq:U_BLR}
U_\mathrm{BLR} = \frac{L_\mathrm{BLR}}{\pi c R_\mathrm{BLR}^2}\,, 
\end{equation}
while the value of $\delta_\mathrm{BLR}$ is obtained considering an isotropic photon field in the LF such that $\langle\cos{\theta_\mathrm{BLR}}\rangle = 0$, and for simplicity is mono-directional in the CF. Under these considerations, for $z < R_\mathrm{BLR}$, $\delta_\mathrm{BLR}= \Gamma_\mathrm{c}$. In the monochromatic approximation, the number of target BLR photons is $n_\mathrm{BLR} = U_\mathrm{BLR}/\epsilon_\mathrm{BLR}$, where $\epsilon_\mathrm{BLR}$ is the characteristic BLR photon energy and transforms as $\epsilon'_\mathrm{BLR} = \epsilon_\mathrm{BLR} \delta_\mathrm{BLR}$. 

To test whether SSC losses are relevant, we estimate the energy density of the synchrotron photon field (calculated neglecting SSC) as $U'_\mathrm{sy} = L'_\mathrm{sy}/(\pi c R_\mathrm{c}^2)$. 
 

\subsection{Particle energy distribution}
   
We consider that a fraction $\xi_\mathrm{e} = 0.1$ of the available injected energy in the shock goes into accelerating relativistic electrons: $L'_\mathrm{inj,e}=\xi_\mathrm{e} L'_\mathrm{inj}$. We adopt a phenomenological injection function for electrons $Q'(E') = K {E'}^{-\alpha} \exp{-E'/{E'}_\mathrm{max}}$, with $\alpha = 2$, and the normalisation constant given by $\int Q'(\tilde{E}') \tilde{E}' \diff \tilde{E}' = L'_\mathrm{inj,e}$. The particle acceleration and cooling times are much shorter than the dynamical timescales, and therefore the particle energy distribution reaches a steady state before the cooling conditions have changed significantly. The solution for the transport equation is 
\begin{equation}
\label{eq:Ne_norm}
 N'_\mathrm{e}(E') = \dot{E'}^{-1} \int_{E'}^{E'_\mathrm{max}} Q'_\mathrm{e}(\tilde{E}') \diff \tilde{E}'\,,
\end{equation} 
with $\dot{E'} = E'/t'_\mathrm{cool}$. The results of the model, in particular at HE, are not very sensitive to $\alpha$ unless it is well above 2, meaning little luminosity in the $\gamma$-ray band. We take a minimum electron energy at injection 
of $E'_\mathrm{min,inj} \sim 1$~MeV, although this parameter does not  significantly affect our results unless it is $\gtrsim 1$~GeV (see  Sect.~\ref{sec:free_parameters}).


We calculate the IC cooling with the parametrisation given by \cite{Khangulyan2014} for an isotropic radiation field. We define the temperature of the radiation field in the CF as $T'_\mathrm{BLR} = T_\mathrm{BLR} \delta_\mathrm{BLR}$, and introduce the dilution factor $\kappa_\mathrm{BLR} = L_\mathrm{BLR} {\delta_\mathrm{BLR}}^2 / \left(4 \pi \sigma_\mathrm{SB} R_\mathrm{BLR}^2 T_\mathrm{BLR}^4 \right)$, with $\sigma_\mathrm{SB}$ the Stefan-Boltzmann constant. We define the constant $C = \pi \hbar^3/(2 r_\mathrm{e}^2 m_\mathrm{e}^3 c^4)$, a normalised temperature $\tilde{T} = k\,T/(m_\mathrm{e} c^2)$, and the parameter $C_\mathrm{BLR} = C/(\kappa_\mathrm{BLR} {\tilde{T'}_\mathrm{BLR}}^2)$. We calculate the IC loss time with the BLR field as 
$t'_\mathrm{IC} = C_\mathrm{BLR} \, \gamma_\mathrm{e} /F_\mathrm{iso}(u'_\mathrm{BLR})$, with $u'_\mathrm{BLR} = 4 \gamma_\mathrm{e} \tilde{T'}_\mathrm{BLR}$ and
\begin{equation}
F_\mathrm{iso}(u) = \left[ \frac{5.68 \, u \ln{\left(1 + \frac{0.722 u}{5.68} \right)}}{1+\frac{5.68 u}{0.822}} \right]
\left[ 1 + \frac{-0.362 \, u^{0.682}}{1 + 0.826 \, u^{1.281}} \right]^{-1}\,.
\end{equation}

The magnetic field in the shocked region is assumed to be isotropic, so that the perpendicular component of the magnetic field is $B'_\perp = \sqrt{2/3} B'$ (this is also taken into account when computing the synchrotron emissivity). The synchrotron losses are given by $t'_\mathrm{sy} = (1.6 \times 10^{-3} B^2 E')^{-1}$~s, which is accurate for an isotropic magnetic field. 

The NT particles are convected by the jet material that flows along the bow-shock. For the convection and adiabatic losses we make the following consideration. As we are working under the one-zone model approximation, the relevant characteristic size in which the target fields change is of the order of $z_\mathrm{c}$. 
We thus take $t'_\mathrm{conv/ad} = z_\mathrm{c} /\left( v'_\mathrm{j} \, \Gamma_\mathrm{c} \right)$, where $v'_\mathrm{j} = (v_\mathrm{j} - v_\mathrm{c})/(1-\beta_\mathrm{j} \, \beta_\mathrm{c})$.
The radiative losses dominate over the non-radiative losses if ${t'}_\mathrm{r} < {t'}_\mathrm{nr}$, with $t'_\mathrm{r} = ({t'}_\mathrm{sy}^{-1} + {t'}_\mathrm{IC}^{-1})^{-1}$ and ${t'}_\mathrm{nr} = {t'}_\mathrm{conv/ad}$. The cooling time is then $t'_\mathrm{cool} = \min{(t'_\mathrm{r}, t'_\mathrm{nr})}$.

The acceleration time for relativistic electrons is $t'_\mathrm{acc} = \eta_\mathrm{acc} E'/(B'_\perp \, c \, q_\mathrm{e})$, where $q_\mathrm{e}$ is the elementary charge, and we fix the acceleration efficiency to $\eta_\mathrm{acc} = 0.1$. The electron maximum energy is obtained as the minimum from the condition $t'_\mathrm{acc}(E'_\mathrm{max})=t'_\mathrm{cool}(E'_\mathrm{max})$ and accounting for diffusive losses.


\subsection{Non-thermal emission}
 
We estimate the IC emission considering a monochromatic, homogeneous, and isotropic BLR photon field. In the CF, the electrons `see' that the BLR photons come from a direction given by $\langle \cos{\theta'_\mathrm{BLR}}\rangle = -\beta_\mathrm{c}$. During the early stages of the JCI, $\Gamma_\mathrm{c} \sim 1$ and the BLR photon phield for the relativistic electrons is nearly isotropic, but for later stages with $\Gamma > 2$ (which are the more relevant ones in the context of this work), this photon field in the CF is nearly mono-directional because of relativistic effects, and the interactions of the relativistic electrons with the BLR photons proceed as (quasi) head-on collisions.

First, we calculate the emitted luminosity in the CF. The BLR photon field is characterised by the photon energy $\epsilon'_\mathrm{BLR}$ and the photon number density $n'_\mathrm{BLR}$ given in \S~\ref{subsec:rad_fields}. In the monochromatic approximation, we consider $\epsilon_\mathrm{BLR} \approx \epsilon_\mathrm{Ly \alpha} \approx 10$~eV. We use the angle-dependent expression for $\sigma_\mathrm{IC}$ \citep[e.g.][]{Khangulyan2014} with an angle $\alpha' = \arccos{(-\beta_\mathrm{c})}$. The SED (i.e. the emitted specific luminosity) in the CF is computed as
\begin{equation}
L'_\mathrm{IC}(\epsilon') = \epsilon' \int_{\epsilon'}^{E'_\mathrm{max}} N'_\mathrm{e}(\tilde{E'}) \, c \, n'_\mathrm{BLR} \sigma_\mathrm{IC}(\alpha',\epsilon',\tilde{E'}) \diff \tilde{E'}\,.
\end{equation}
The total luminosity at a given time is $L' = \int L'(\epsilon') \diff \epsilon'$. We also compute the synchrotron luminosity in the CF from the particle energy distribution. However, we do not study the synchrotron component in detail as it is very sensitive to the adopted values for $\eta_B$ and $E'_\mathrm{min,inj}$ (see forthcoming Sect.~\ref{sec:free_parameters}). We also note that we do not expect the JCIs to dominate the synchrotron emission in blazars, but instead to be a minor contribution to the overall observed flux. To properly model the synchrotron SED, additional absorption processes such as synchrotron self-absorption should be considered in Sect.~\ref{sec:absorption}.
Second, we calculate the expected luminosity in the OF, which is related to the emitted luminosity in the CF as $\epsilon \hat{L}(\epsilon)=\delta_\mathrm{c}^4 \epsilon'L'(\epsilon')$, where 
$\delta_\mathrm{c}~=~\left[(1-\beta_\mathrm{c} \cos{\theta_\mathrm{obs}})\Gamma_\mathrm{c}\right]^{-1}$ is the Doppler boosting factor.


\subsection{Absorption} \label{sec:absorption}

The $\gamma$-ray photons emitted in the shock have to escape the AGN before reaching the observer. During their propagation, these photons can interact with ambient photons 
producing $e^{\pm}$ pairs, resulting in their annihilation. This absorption process is more effective when the photons interact head-on and 
if the ambient photon field intensity is high. As shown by several authors \citep{Liu2006, Finke2016, Bottcher2016, Abolmasov2017}, $\gamma$ rays with 
energies between 30~GeV and $\sim$~TeV emitted within $R_\mathrm{BLR}$ are completely absorbed in the BLR photon field; moreover, the torus radiation
field is totally opaque for $\gamma$ rays with energies exceeding a few TeV
for even larger distances \citep[]{Donea2003}. 
Therefore, to the first order, we can consider the $\gamma$-ray emission to be unabsorbed for $ \epsilon \lesssim 30$~GeV and totally absorbed for $\epsilon \gtrsim 30$~GeV. At this point we do not calculate the radiation from electromagnetic cascades that could potentially result in a more transparent medium for $\gamma$-ray photons. However, we can qualitatively assess the importance of the emission coming from secondary pairs.

To test whether pair emission could be relevant, we need to compare the predicted OF $\gamma$-ray luminosity in the $0.1-30$~GeV energy band, $\hat{L}_{0.1-30}$, to the OF luminosity produced by the secondary pairs. If these pairs are created within the jet, they will be attached to the jet flow and their emission luminosity ($L_{e^\pm}$) beamed; under fast cooling these pairs will have an OF luminosity comparable to that absorbed (as seen by the observer). Within the jet, it is possible that the boosted BLR field will dominate over the magnetic field, and IC cascades could develop. If the pairs isotropise in the LF outside the jet, then their luminosity towards the observer will be $\sim 1/2\Gamma_{\rm j}^2$ times lower. In the LF outside the jet, the magnetic field may be more relevant, in which case pairs would cool through synchrotron. The maximum value of the luminosity of the pairs $\hat{L}_{e^\pm}$ depends on the emitted (i.e. unabsorbed) and observed (i.e. absorption corrected) luminosities as
$\Delta \hat{L} = \hat{L}_\mathrm{\gamma,em} - \hat{L}_{\gamma} \sim \hat{L}_\mathrm{>30GeV,em}$, and is between  
$\sim\Delta \hat{L}/2\Gamma_{\rm j}^2$ and $\Delta \hat{L}$.


\section{Results of a typical interaction} \label{sec:results}

\begin{figure}
   \resizebox{\hsize}{!}{\includegraphics[width=0.65\linewidth, angle=270]{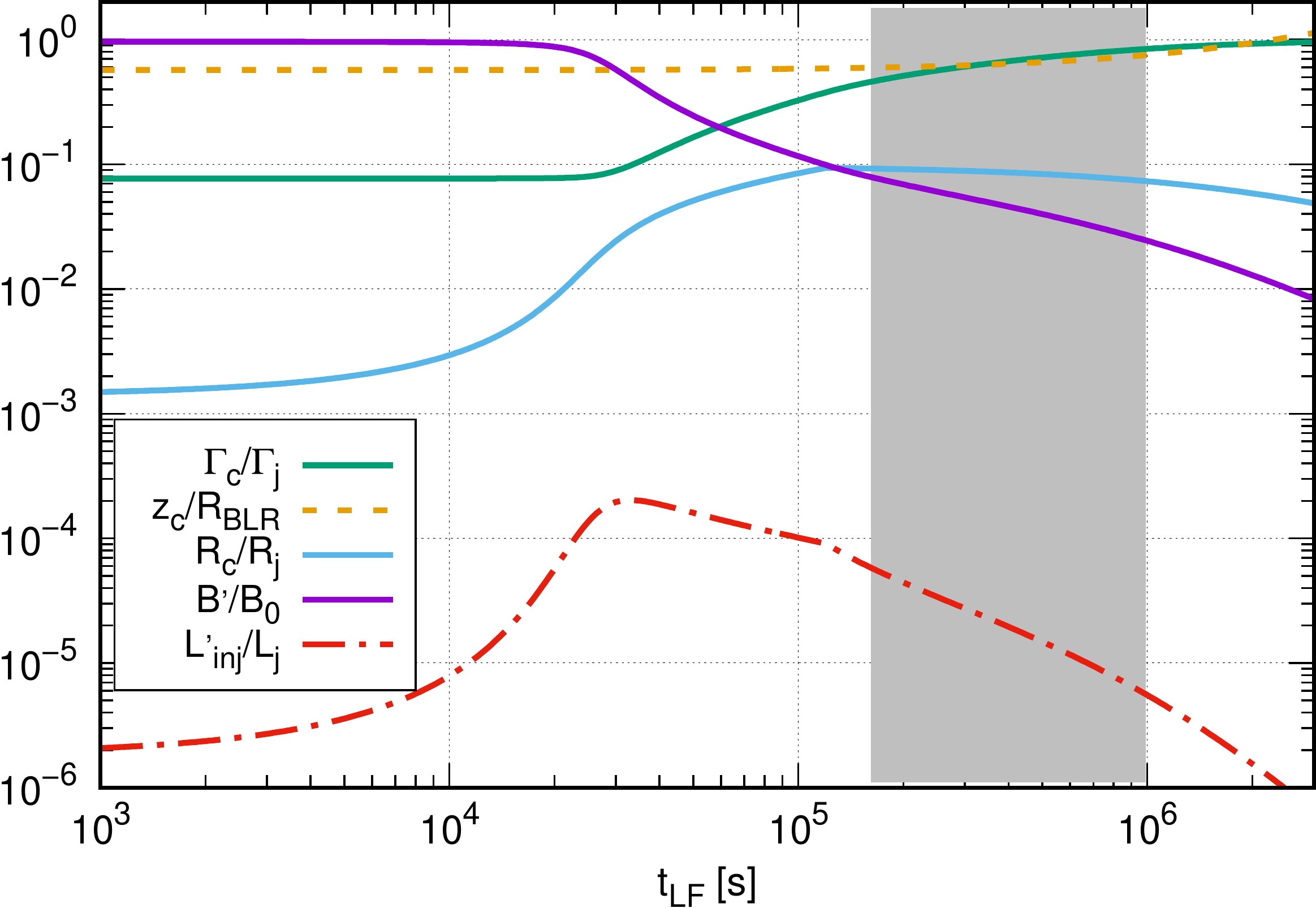}}
   \caption[]{Evolution of the dynamical quantities in a typical JCI. The grey shaded area represents the radiatively efficient stage in the OF.}
   \label{fig:dynamic_norm}
 \end{figure}

We present the results of the interaction of a single JCI under typical conditions (parameters given in Table~\ref{tab:parameters}). From Eq.~\ref{eq:z_min} we obtain $z_\mathrm{min} \approx 5\times 10^{16}$~cm and $z_\mathrm{j} \approx 10^{17}$~cm. Equation~\ref{eq:B} yields $B_0 \approx 350$~G. The evolution of the relevant dynamical quantities is shown in Fig.~\ref{fig:dynamic_norm}. In the LF, after $\sim 10^4$~s the cloud expands considerably, up to $\sim 10\%$ of the jet radius, and after $\sim 4\times 10^4$~s it accelerates to a Lorentz factor $> 2$, reaching $\Gamma_\mathrm{c} \approx \Gamma_\mathrm{j}$ after $\sim 10^6$~s. The magnetic field in the CF drops significantly as $\Gamma_\mathrm{c}$ increases (from Eq.~\ref{eq:B}, $B' \lesssim B/\Gamma_\mathrm{c}$). 
During the early, non-relativistic stage ($t_\mathrm{LF} < 10^4$~s), the momentum transfer from the jet to the cloud is significant, but not the energy transfer. 
Later, in the relativistic stage ($t_\mathrm{LF} > 10^4$~s), the transfer of energy from the jet to the cloud (in the CF) becomes more efficient. Once the cloud accelerates and $z_\mathrm{c}$ starts growing significantly ($t \gtrsim 10^5$~s), the cloud expansion is negligible and the fraction of the jet section it occupies becomes smaller (Fig.~\ref{fig:dynamic_norm}). When $\Gamma_\mathrm{c} \rightarrow \Gamma_\mathrm{j}$, the relative velocity between the jet and the shock becomes very small and so does the energy flux through the shock, $q'$ (Eq.~\ref{eq:P_j};  the available luminosity per unit surface also diminishes); in consequence, no further energy is deposited in NT particles.
   
The relativistic particle energy distribution normalisation depends on $L'_\mathrm{inj}$, which varies as a function of time (see Fig.~\ref{fig:dynamic_norm}). Qualitatively, $L'_\mathrm{inj}$ increases as the cloud expands and the shocked jet area becomes larger, but once the cloud has accelerated significantly $q'$ decreases and so does $L'_\mathrm{inj}$. 
\begin{figure}
 \resizebox{\hsize}{!}{\includegraphics[angle=270]{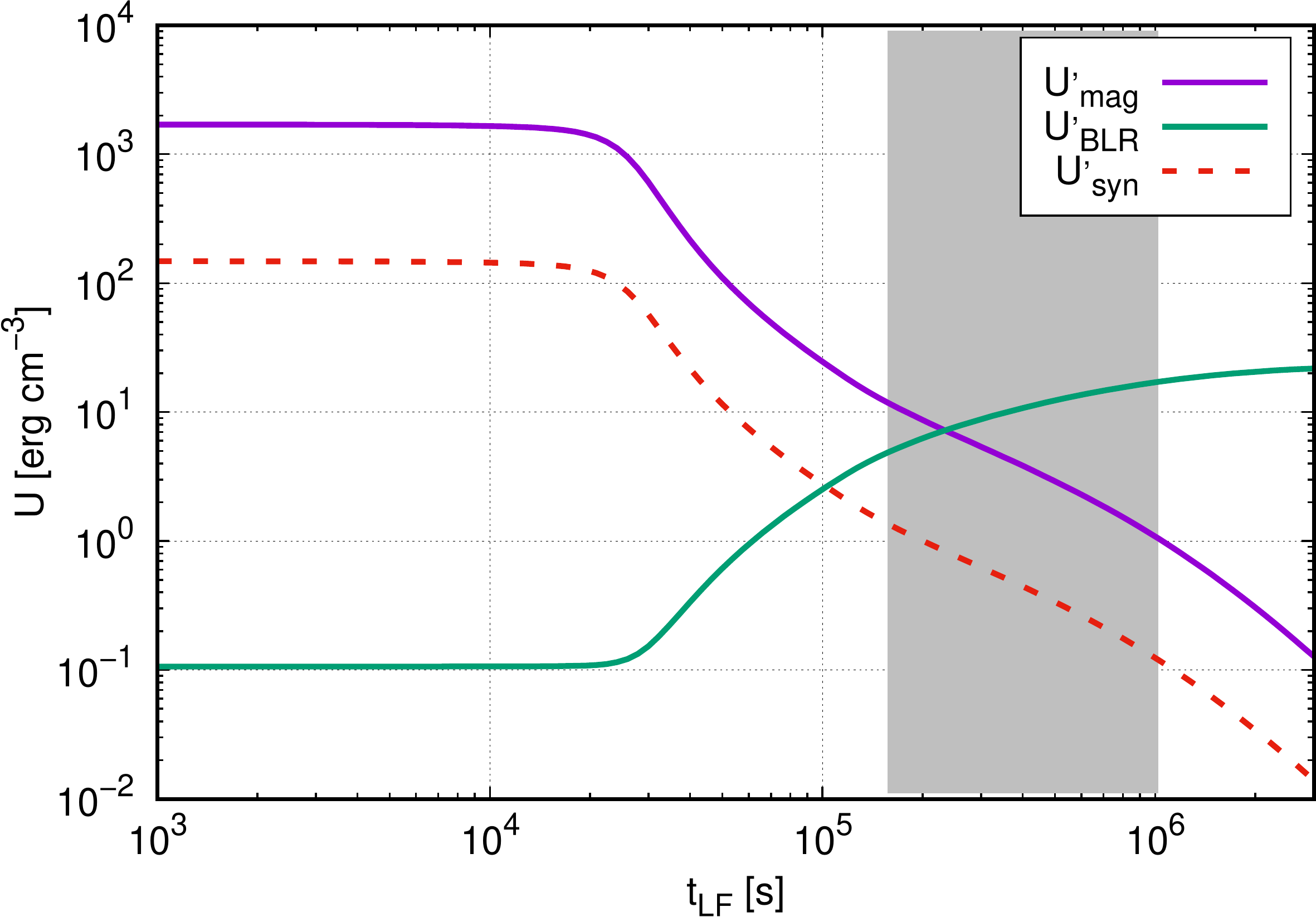}}
 \caption[]{Evolution of different energy densities in the CF during a JCI. For $t_\mathrm{LF} \lesssim 2\times 10^5$~s the synchrotron cooling is dominant (followed by SSC, not considered in the model). For $t_\mathrm{LF} \gtrsim 2\times10^5$~s the BLR photon field dominates the cooling of electrons that interact in the Thomson regime. The grey shaded area represents the radiatively efficient stage in the OF.}
 \label{fig:U_vs_t}
\end{figure}

\begin{figure}
 \resizebox{\hsize}{!}{\includegraphics[angle=270]{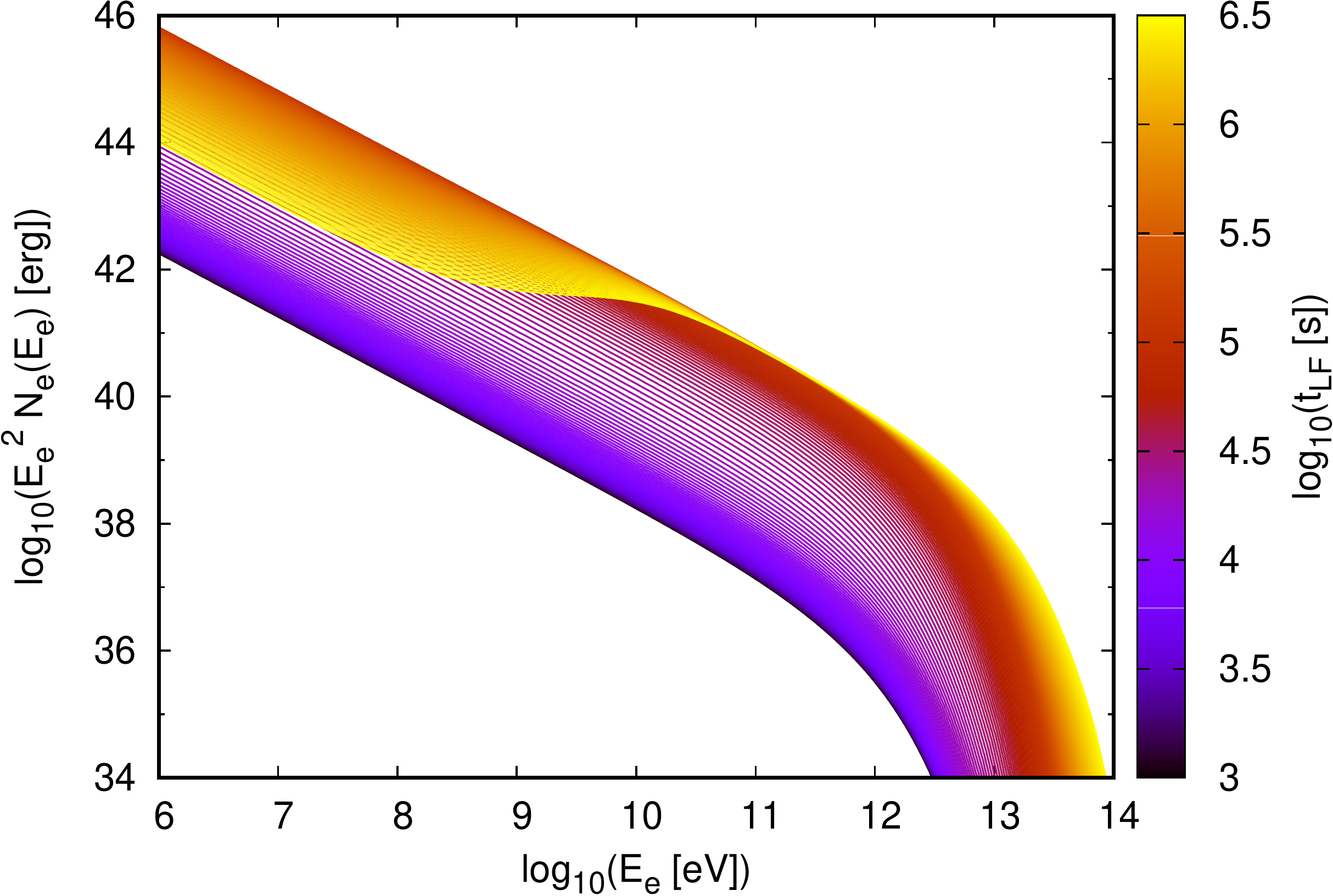}}
 \caption[]{Relativistic electron energy distribution for different times in the LF during the JCI. The distribution is roughly $N'_\mathrm{e}(E') \propto {E'}^{-3}$, as expected when IC-Thomson or synchrotron losses are dominant. At $t_\mathrm{LF} \sim 10^6$~s, the IC cooling for electrons with $E' \sim 10^9 - 10^{10}$~eV occurs in the K-N regime, which leads to a hardening in the electron distribution up to energies at which synchrotron cooling takes over and softens the particle distribution.}
 \label{fig:dist_vs_t}
\end{figure}

The shape of the relativistic particle energy distribution depends on the dominant cooling/escape mechanisms. The IC/SSC and synchrotron losses depend on the energy density of the target photon fields and the magnetic field, respectively (see Fig.~\ref{fig:U_vs_t}). The IC losses also depend on the product of the electron and target photon energies: if $E' \, \epsilon' > (m_\mathrm{e} c^2)^2$, the interaction occurs in the Klein-Nishina (K-N) regime, which has a much smaller interaction probability, making the IC cooling less efficient. 
For early times, the dominant cooling mechanism is synchrotron, whereas at later times (which are  more relevant in terms of $\gamma$-ray emission) the cooling is dominated by IC scattering with the boosted BLR photon field. However, electrons with $E' \gtrsim 1-10$~GeV interact in the K-N regime once the cloud reaches $\Gamma_\mathrm{c} \gg 1$ \footnote{As $\epsilon'_\mathrm{BLR} \sim 10 \, \Gamma_\mathrm{c}$~eV, the transition to the KN regime occurs at $E'_\mathrm{KN} \approx 25 \, \Gamma_\mathrm{c}^{-1}$~GeV. The electrons interact with the torus photon field in the Thomson regime to significantly higher energies, which can become a relevant cooling mechanism for electrons with $E' \gtrsim 100$~GeV once the cloud has reached $\Gamma_\mathrm{c} \gtrsim 10$. However, this process does not dominate the SED in the GeV energy range, and is therefore neglected for simplicity.}.
The electron maximum energy is limited by synchrotron cooling, so it increases as a function of time as $B'$ decreases. The electron maximum energy obtained is $\approx 1$~TeV at early stages, and it later increases up to $\sim 10$~TeV. 
Electrons cool down locally regardless of their energy. The radiatively efficient stage is during $t_\mathrm{LF} \sim 10^5-10^6$~s, and is shown as a grey shaded area in Figs.~\ref{fig:dynamic_norm}--\ref{fig:U_vs_t}. The electron energy distribution for different times (in the LF) is shown in Fig.~\ref{fig:dist_vs_t}. The distribution is a soft power law with $N'_\mathrm{e}(E') \propto {E'}^{-3}$, except for $t_\mathrm{LF} \sim 10^6$ when a hardening for electrons with energies $E' \sim 10^9 - 10^{10}$~eV appears because IC is in the K-N regime; the distribution softens again at higher energies as synchrotron cooling takes over. 
%
  
\begin{figure}
 \resizebox{\hsize}{!}{\includegraphics[]{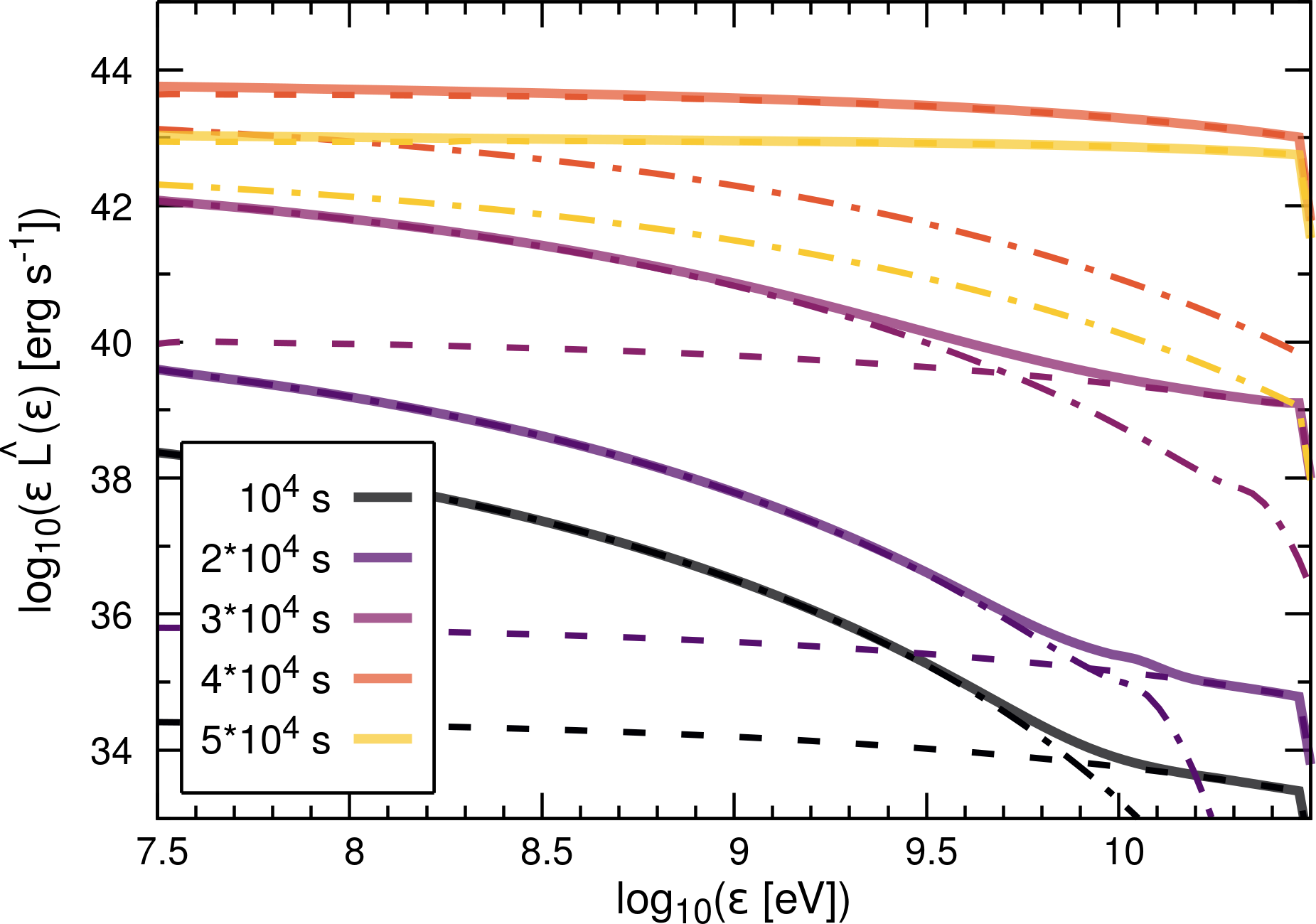}}
 \caption[]{SEDs for different observing times. The dashed and dash-dotted lines represent the IC and synchrotron contribution, respectively, while the solid lines are the total emission. The emission increases with time until $4\times 10^4$~s, when it starts to decrease. The synchrotron emission dominates at energies $< 10$~GeV for early times, but IC dominates the whole SED above $0.1$~GeV during the most luminous stages.}
 \label{fig:SED_vs_tobs}
\end{figure}

\begin{figure}
 \resizebox{\hsize}{!}{\includegraphics[angle=270]{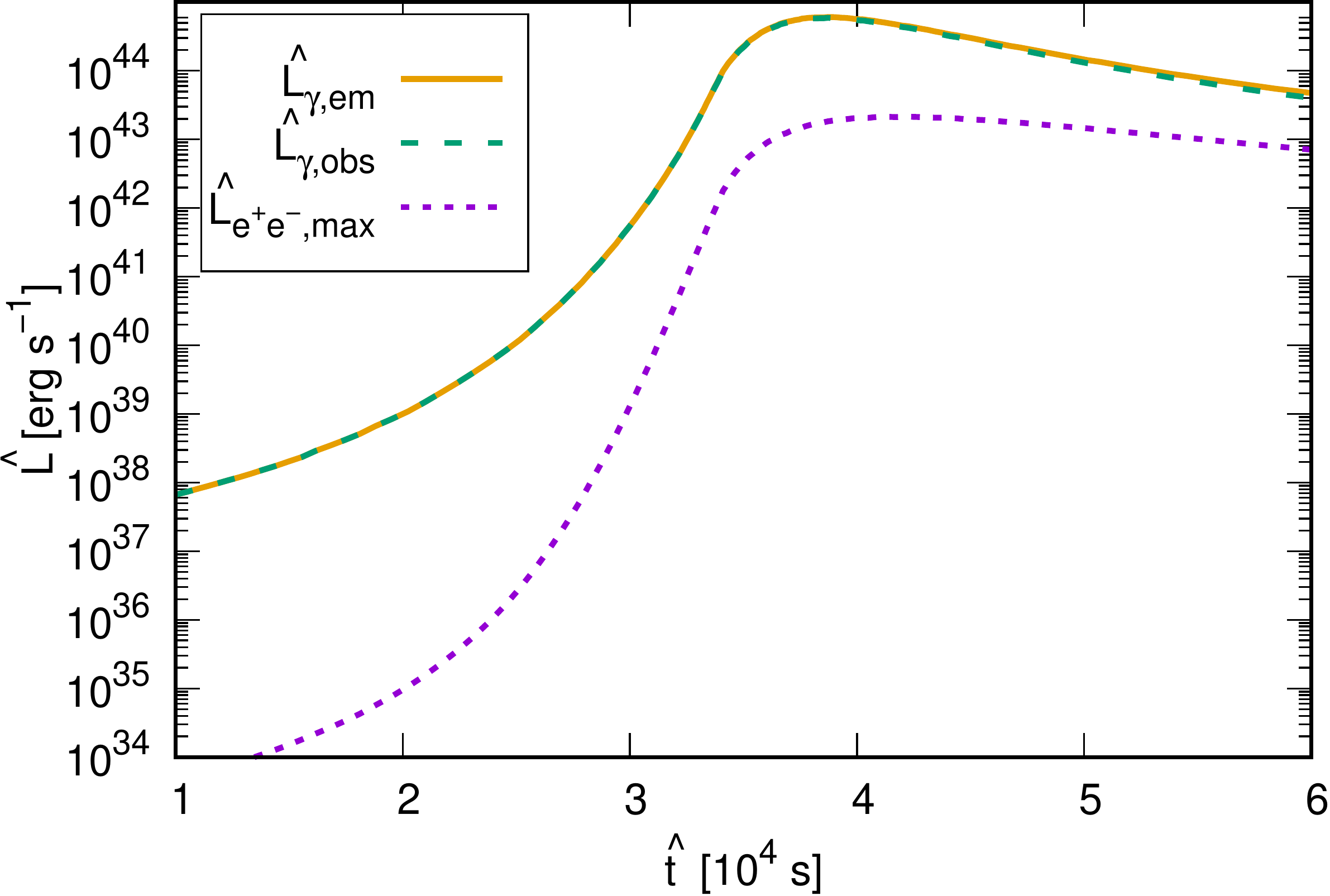}}
 \caption[]{Comparison between the emitted (i.e. without absorption) and the observed (i.e. absorption corrected) integrated $\gamma$-ray luminosity above $0.1$~GeV in the OF, $\hat{L}_{\gamma,\mathrm{em}}$ and $\hat{L}_{\gamma,\mathrm{obs}}$, respectively, and the maximum luminosity produced by pairs, $\hat{L}_{e^\pm,max}$, estimated as $\hat{L}_{e^\pm,\mathrm{max}} = \hat{L}_{\gamma,\mathrm{em}} - \hat{L}_{\gamma,\mathrm{obs}}$.}
 \label{fig:Labs_vs_Linj}
\end{figure}

In Fig.~\ref{fig:SED_vs_tobs} we show the evolution of the synchrotron and IC-BLR SEDs for different observing times. 
During the early stage, the electrons cool down more efficiently through synchrotron cooling, although this process hardly generates emission above 0.1~GeV. However, as the interaction evolves, the shocked cloud accelerates from $\Gamma_\mathrm{c} \approx 1$ to $\Gamma_\mathrm{c} \approx \Gamma_\mathrm{j}=13$. Thus, due to the increased BLR photon field in the CF, for later stages the IC losses with BLR photons dominate the radiative cooling and $\hat{L}_\mathrm{IC}$ becomes larger than $\hat{L}_\mathrm{sy}$. The IC-BLR process is responsible for most of the emission in the $1-30$~GeV energy band. Moreover, Doppler boosting has the effect of displacing the SEDs to higher energies by a factor $\delta_\mathrm{c} \lesssim \Gamma_\mathrm{j}$  and enhancing the flux by a factor $\delta_\mathrm{c}^4 \lesssim \Gamma_\mathrm{j}^4$. Thus, Doppler boosting has a major effect in the shape of the SED, in particular enhancing the emission by more than four orders of magnitude for late stages. The synchrotron process can radiate appreciably in the $0.1-1$~GeV energy range due to the Doppler emission combined with the presence of very high-energy electrons ($\gtrsim 10$~TeV, Fig.~\ref{fig:dist_vs_t}) at late stages, under the adopted acceleration efficiency. In addition, Fig.~\ref{fig:Labs_vs_Linj} shows that the fraction of the luminosity that goes into secondary $e^{\pm}$ pairs is not dominant, although the absorption process is relevant in shaping the IC spectrum at energies $>30$~GeV. 

\begin{figure}
 \resizebox{\hsize}{!}{\includegraphics[angle=270]{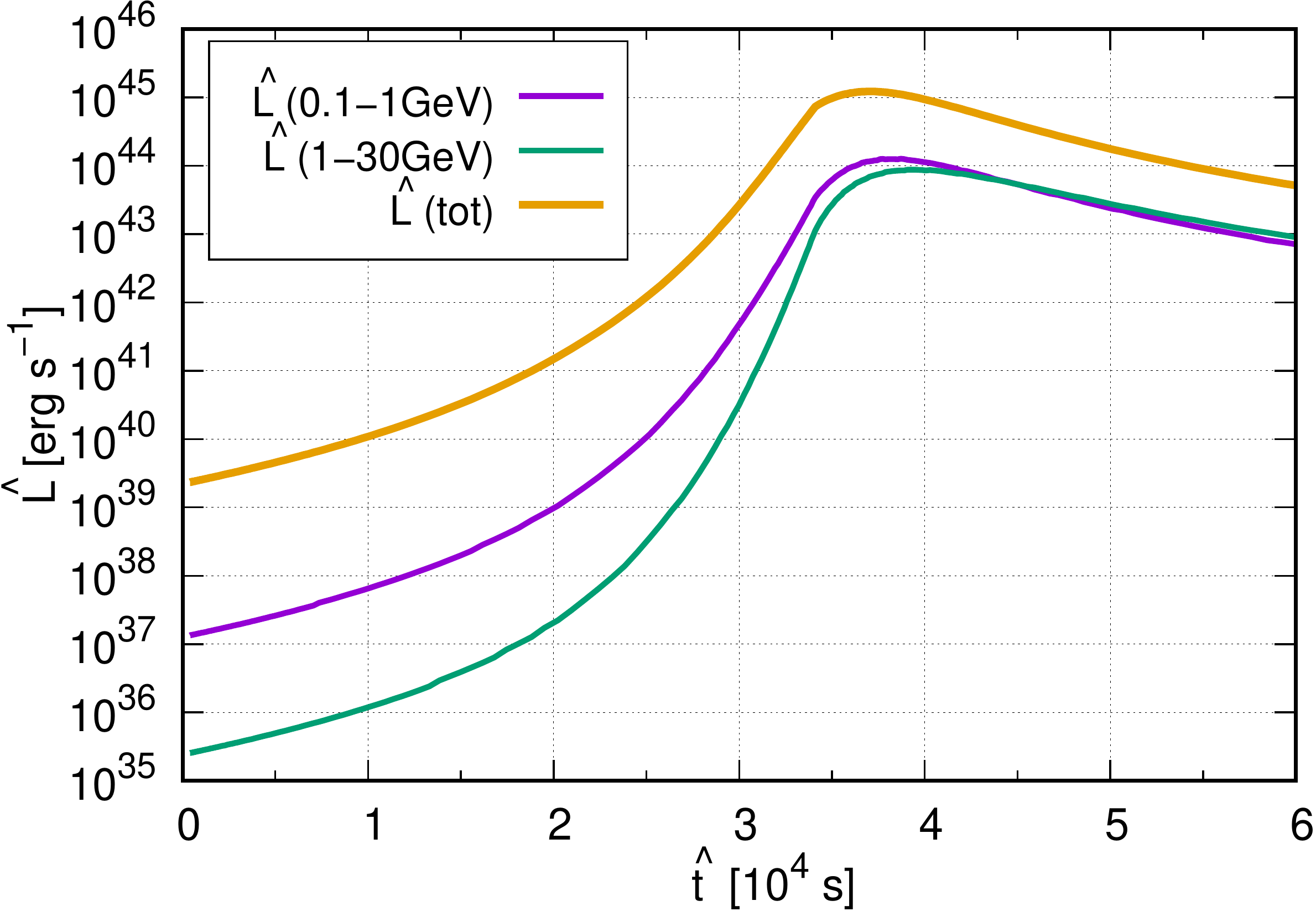}}
 \resizebox{\hsize}{!}{\includegraphics[angle=270]{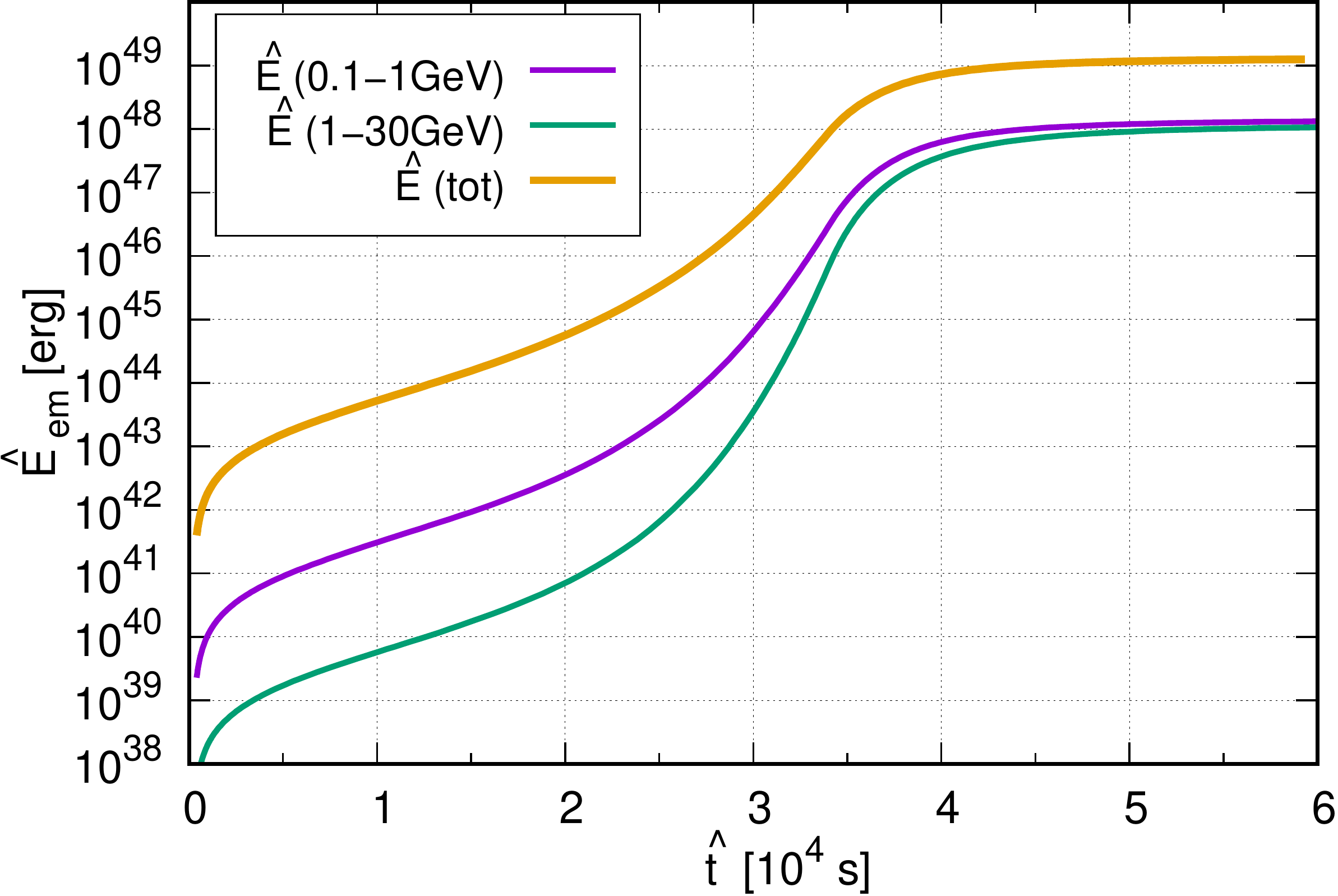}} 
 \caption[]{Luminosity (top) and integrated luminosity (i.e. radiated energy; bottom) in the OF as a function of time, for different energy bands. The majority of the observed emission occurs between $\hat{t} = 4\times 10^4$~s and $\hat{t} = 2\times 10^6$~s, in the range $0.1-30$~GeV. The $\gamma$-$\gamma$ absorption has a dramatic effect on photon energies above $30$~GeV.} 
\label{fig:L_bands_vs_tobs}
\end{figure}

\begin{figure}
 \resizebox{\hsize}{!}{\includegraphics[angle=270]{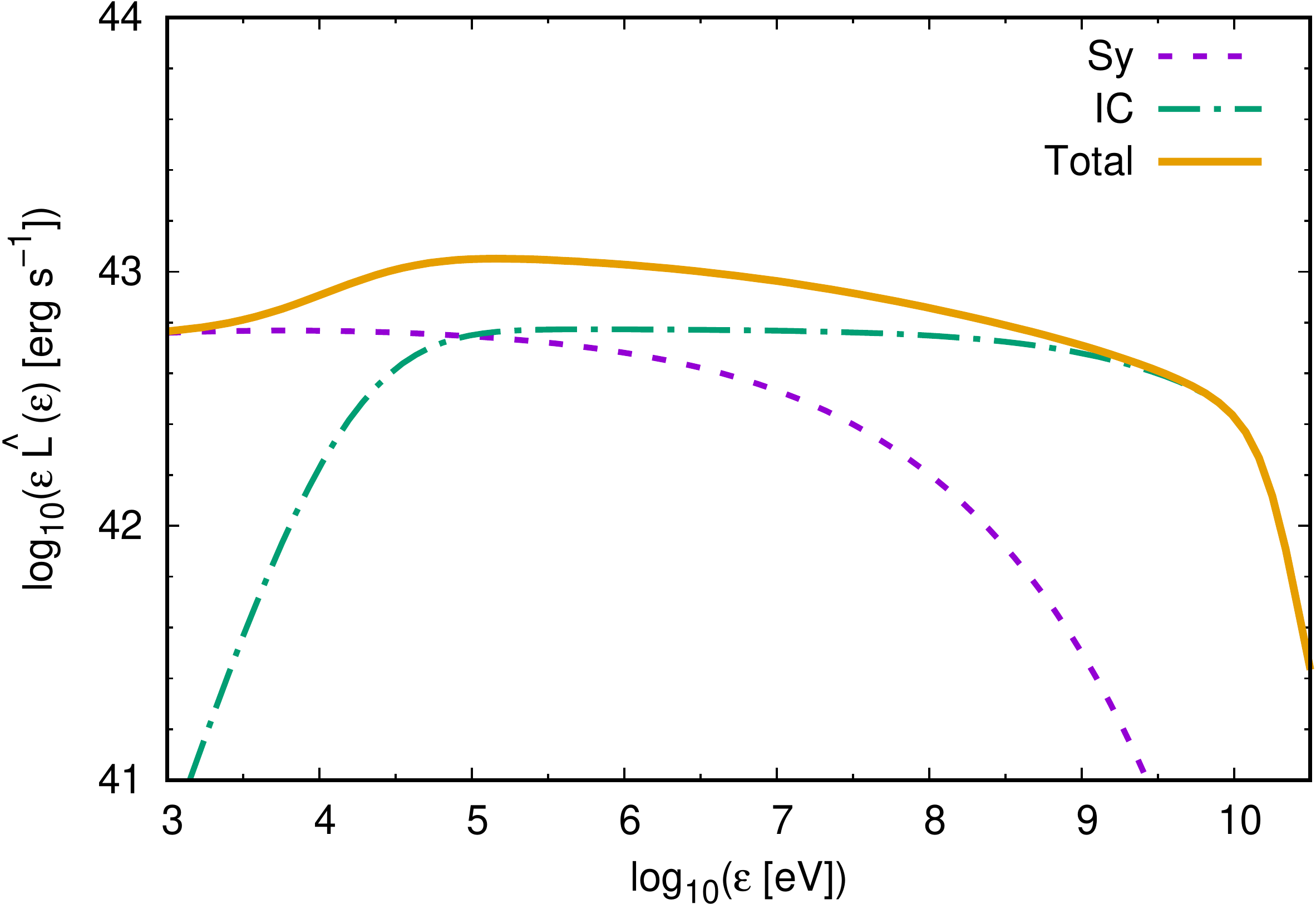}}
 \resizebox{\hsize}{!}{\includegraphics[angle=270]{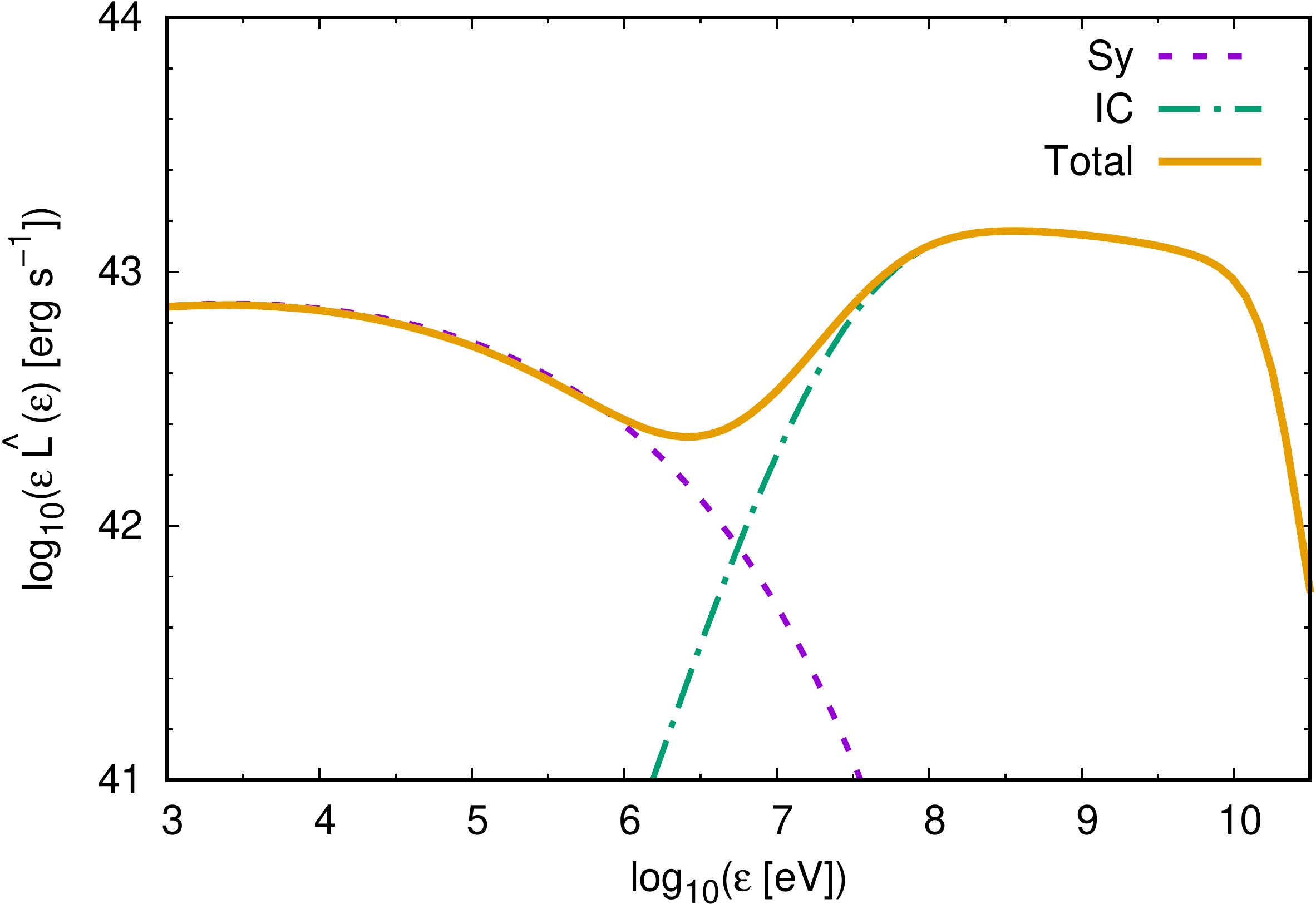}}
 \caption[]{Time-averaged synchrotron and IC SEDs for a typical JCI (top) and for the \textit{\emph{favourable}} scenario described in Sect.~\ref{sec:free_parameters} (bottom).}
 \label{fig:SED_ave}
\end{figure}

The lightcurve for a single JCI has a sharp peak at $\hat{t} \approx 4\times10^4$~s (Fig.~\ref{fig:L_bands_vs_tobs}). In the OF, the radiatively significant stage of the interaction lasts for $\hat{t}\sim 10^4$~s. It is interesting to see whether the emission in the 0.1--30~GeV energy range presents different signatures in sub-bands (Fig.~\ref{fig:L_bands_vs_tobs}). The emission at $\epsilon > 10$~GeV has the most significant variation, as it is produced only by IC scattering with the BLR photons. 
In Fig.~\ref{fig:L_bands_vs_tobs} we also show the integrated luminosity (i.e. total radiated energy) received in different energy bands. The total emitted energy in the 0.1--1~GeV energy range is almost twice the emitted at 1--10~GeV, leading to a SED with an average slope $\alpha \approx -0.2$. This can be explained in terms of the flat IC SED and a significant synchrotron contribution at $\epsilon < 1$~GeV (Fig.~\ref{fig:SED_vs_tobs}). Figure~\ref{fig:SED_ave} shows that the time-averaged SED above $100$~keV is completely dominated by IC with BLR photons. The time-averaged total luminosity is $\langle \hat{L}_\gamma \rangle \approx 1.5 \times 10^{44}$~erg~s$^{-1}$ during each interaction, and in the $0.1-30$~GeV energy band it is $\langle \hat{L}_{0.1-30} \rangle \approx 2.9 \times 10^{43}$~erg~s$^{-1}$. 
%
%

 

\subsection{Emissivity scaling on free parameters} \label{sec:free_parameters}
   
The value of the equipartition magnetic field considered (corresponding to $\eta_B =1$) is rather high, so it is worth  analysing the impact that this assumption could have on the predicted luminosity of a JCI. If the magnetic field is less intense ($\eta_B < 1$), the 
IC-BLR dominates the electron cooling earlier than suggested in Fig.~\ref{fig:U_vs_t}. In consequence, the $\gamma$-ray luminosity peaks slightly sooner and at slightly higher values. 
%
%
The predicted time-averaged $\gamma$-ray luminosity in the 0.1--30~GeV energy range grows from $\langle \hat{L}_{0.1-30} \rangle \approx 3\times 10^{43}$~erg~s$^{-1}$ for $\eta_B = 1$ to $\langle \hat{L}_{0.1-30} \rangle \approx 5\times 10^{43}$~erg~s$^{-1}$ for $\eta_B = 0.1$; 
smaller values of $B$ would be in contradiction with a jet magnetically dominated at its base. In addition, the $\gamma$-ray spectrum becomes slightly harder for smaller $\eta_B$ as the flat IC component becomes more dominant than the softer synchrotron component; 
also the Compton dominance (given by the ratio of the IC and synchrotron fluxes) increases for smaller values of $\eta_B$.
%

Another free parameter of our model is the minimum energy of the injected relativistic electron population, $E'_\mathrm{min,inj}$. Qualitatively, a higher value of $E'_\mathrm{min,inj}$ implies that there is more energy available for higher energy electrons (as can be inferred from Eq.~\ref{eq:Ne_norm}), and also that the low-energy tail of the IC-BLR SED shifts to higher energies (roughly $\epsilon_\gamma \gtrsim E'_\mathrm{min,inj}$). Increasing $E'_\mathrm{min,inj}$ from 1~MeV to 10--50~MeV increases $\langle \hat{L}_{0.1-30} \rangle$ by a  15--30\%.

At last, the acceleration efficiency of relativistic particles is also a free parameter of the model and it affects  the maximum energy achieved by the relativistic electrons. The adopted value of $\eta_\mathrm{acc} = 0.1$ for a typical JCI is somewhat high, but even considering a much smaller value of $\eta_\mathrm{acc} = 10^{-3}$ only has a minor impact in $\langle \hat{L}_{0.1-30} \rangle$, enhancing it by a 5~\%. However, the value of $E'_\mathrm{max}$ can have a significant impact in the shape of the SED as it determinates the maximum photon energy for each process. In the case of IC-BLR, the maximum value of $\epsilon_\gamma$ is hardly relevant as all photons with $\epsilon_\gamma > 30$~GeV are absorbed. For synchrotron radiation, considering a lower value of $\eta_\mathrm{acc}$ shifts the synchrotron peak to lower frequencies. In Fig.~\ref{fig:SED_ave} we also show an example of a SED for a \textit{\emph{favourable}}  scenario with $\eta_B = 0.1$, $\eta_\mathrm{acc} = 10^{-3}$ and $E'_\mathrm{min,inj} = 50$~MeV. In this case the SED has the characteristic two humps  of blazar SEDs and the $\gamma$-ray luminosity is almost a factor of three higher than that obtained for the typical values previously considered, $\langle \hat{L}_{0.1-30} \rangle \approx 8\times 10^{43}$~erg~s$^{-1}$.

In summary, the different free parameters in the model ($\eta_B$, $E'_\mathrm{min,inj}$, $\eta_\mathrm{acc}$) can have a strong impact on shaping the broad-band SED of a JCI. Nonetheless, the emitted $\gamma$-ray luminosity in the GeV band by a JCI is not very sensitive to variations in these parameters and the values reported for a typical JCI should be correct within a factor of $\sim 3$. This supports the decision of focusing on the study of the GeV $\gamma$-ray luminosity in JCIs as the results obtained are rather robust and not strongly model-dependent.


\section{Interaction rates} \label{sec:interactions}

To determine whether JCIs give rise to transient/flaring or steady emission, we need to take into account the rate of events and their duration. This can be achieved by estimating the duty-cycle defined as $DC = \dot{N} \, \hat{t}_\mathrm{int}$, where $\dot{N}$ is the number of clouds entering the jet per time unit, and $\hat{t}_\mathrm{int}$ is the duration of each individual interaction in the OF (i.e. the time during which the JCI is visible in the OF). 
We estimate $\dot{N} \sim N_\mathrm{c,j} / t_\mathrm{j}$, with $N_\mathrm{c,j}$ being the number of clouds inside the jet and $t_\mathrm{j}$ the time it would take for a cloud to cross the jet width if there were no interaction with the jet. To estimate this timescale we need to consider a representative value of the cloud velocity, $\langle v_\mathrm{c,0} \rangle$, such that $t_\mathrm{j} \sim  2 R_\mathrm{j}/\langle v_\mathrm{c,0} \rangle$. According to Eq.~\ref{eq:z_min}, the minimum velocity required for a cloud to enter the jet is $v_\mathrm{c,min}\approx 3 \times 10^8$~cm~s$^{-1}$ at $z \sim R_\mathrm{BLR}$ under the assumed conditions. Further assuming that the BLR clouds follow a Maxwellian velocity distribution $g(v_\mathrm{c,0})$ with mean value $\bar{v}_\mathrm{c,0} = \sqrt{GM/R_\mathrm{BLR}}$, the value of the most likely velocity can be estimated as 
\begin{equation}
 \langle v_\mathrm{c,0} \rangle = 
    \int_{v_\mathrm{c,min}}^{\infty} v_\mathrm{c,0} \, g(v_\mathrm{c,0}) \, \mathrm{d}v_\mathrm{c,0}.
\end{equation}
We obtain a typical value of $\langle v_\mathrm{c,0} \rangle \approx 8 \times 10^8$~cm~s$^{-1}$.

We consider the simplest scenario of a spherical BLR with clumps of equal size, and use the canonical values given in Table~\ref{tab:parameters}. The total number of clouds in the BLR is $N_\mathrm{c,tot}$. If we adopt a fixed filling factor $f \sim 10^{-6}$, we can estimate this number as $N_\mathrm{c,tot} \sim f \, (R_\mathrm{BLR}/R_\mathrm{c,0})^3$ \cite[e.g.][]{Araudo2010}. Another possibility is to consider a fixed covering factor $C_\mathrm{BLR} \sim 0.1$, in which case $f = 4/3 (R_\mathrm{c,0}/R_\mathrm{BLR}) C_\mathrm{BLR}$ and $N_\mathrm{c,tot} \sim C_\mathrm{BLR}^3 f^{-2}$ \cite[e.g.][]{Abolmasov2017}; this is the approach followed here, although the difference between the two methods is small. 
The fraction of clouds that have enough velocity to penetrate into the jet (Sect.~\ref{sec:dynamics}) is $f_\mathrm{vel}$, and the total number of clouds inside the jet at a given time is
\begin{equation}
N_\mathrm{c,j} = N_\mathrm{c,tot} \, \left( \frac{\Delta \Omega_\mathrm{j}}{4\pi}  \right)\,, 
\end{equation}
where $\Delta \Omega/(4\pi) = (1-\cos{\theta_\mathrm{j}})/2 \approx (\theta_\mathrm{j}/2)^2 \approx 0.0025$ is the jet-to-BLR volume ratio.

In the OF, the interaction lasts $\hat{t}_\mathrm{int} \sim 5 \times 10^4$~s. 
For $C_\mathrm{BLR} = 0.1$ \citep[e.g.][]{Abolmasov2017}, we obtain $f \approx 8\times10^{-6}$, $N_\mathrm{c,tot} \sim 10^7$, and $DC \sim 10-100$ for typical parameters. Therefore, the emission we expect from JCI events is steady. Considering Poisson statistics, the typical variability timescale we expect is $\hat{t}_\mathrm{int}/DC \sim 10^3\,(DC/30)^{-1}$~s, and fluctuations in luminosity of $\sim DC^{1/2}/DC \approx 0.20\,(DC/30)^{-1/2} \approx 20\,(D/30)^{-1/2}$~\% \citep[e.g.][for a similar approach in a microquasar scenario]{Owocki2009}. This effect is observed in the radio and optical bands \citep{Romero2002}.
The expected luminosity is $DC$ times the luminosity of a single JCI, i.e. $\sim 3\times 10^{45}(DC/30)$~erg~s$^{-1}$. This is similar to the average emission detected from blazars with jet power $L_j \sim 3 \times 10^{46}~\mathrm{erg~s}^{-1}$ \citep{Ghisellini2017}. We can also estimate the average luminosity as $\dot{N}<E_\gamma>\sim 10^{46}\,(\do{N}/10^{-3})(<E_\gamma>/10^{49})$~erg~s$^{-1}$, which yields similar results.

An interesting feature related to the multiple JCIs is the jet mass-loading produced by the BLR cloud entrainment. The mass-loading rate is of the order of $\dot{M}=\dot{N} M_\mathrm{c}\sim 10^{23}\,(\do{N}/10^{-3})(M_{\rm c}/10^{26})$~g~s$^{-1}$. This process is not dynamically relevant for typical values; for instance, under the conditions assumed here, $\dot{M}\ll \dot{M}_{\rm j}\sim L_\mathrm{j}/\Gamma_\mathrm{j} c^2 \sim 10^{24}\,(L_{\rm j}/10^{46})(\Gamma_{\rm j}/10)^{-1}$~g~s$^{-1}$. 


\subsection{Emissivity scaling on jet power}
   
\begin{figure}
 \resizebox{\hsize}{!}{\includegraphics[angle=270]{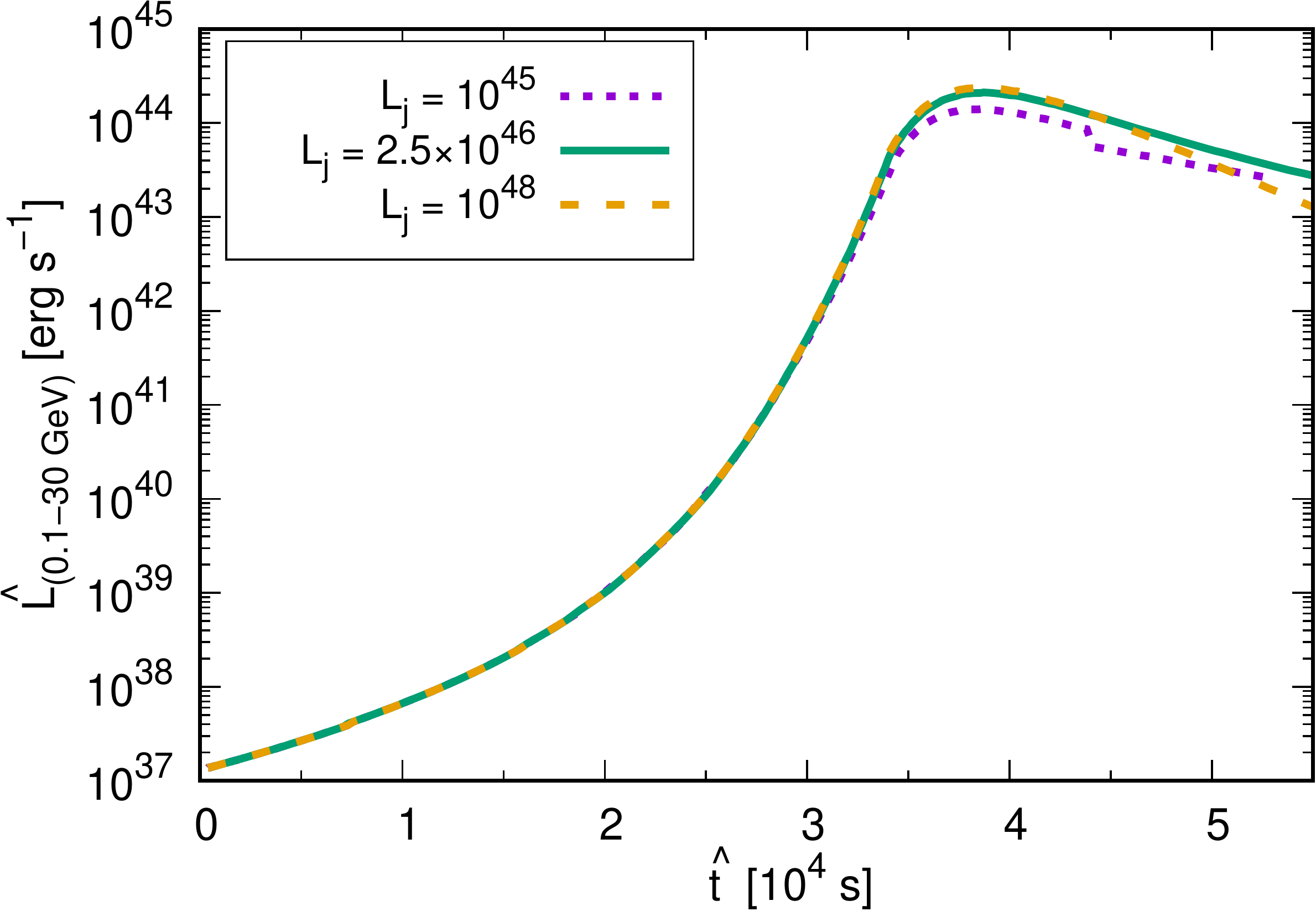}}
 \caption[]{Evolution of the integrated $\gamma$-ray luminosity in the 0.1--30~GeV energy range in the OF for different values of the jet power.}
 \label{fig:L_vs_Lj}
\end{figure}

We compute a typical JCI for the same cloud parameters ($R_\mathrm{c,0}$, $n_\mathrm{c,0}$, and $v_\mathrm{c,0}$) as in Table~\ref{tab:parameters}, but for jet luminosities different from the canonical value ($L_\mathrm{j} \sim 2.5 \times 10^{46}$~erg~s$^{-1}$). 
Regardless of the jet power, the basic characteristics of the JCI are essentially the same: the cloud gets accelerated in $t_\mathrm{LF} \sim 10^6$~s, the high-energy electrons cool down locally through IC interactions with the BLR photon field during the most relevant stage, and the radiatively efficient phase lasts $\sim (1-2)\times 10^4$~s in the OF (Fig.~\ref{fig:L_vs_Lj}). 

For a less powerful jet with $L_\mathrm{j} \sim 10^{45}$~erg~s$^{-1}$, the BLR region is smaller and $D \lesssim 1$. Thus, the cloud escapes faster from this region and the assumption of homogeneous conditions is only valid for penetration heights considerably below $R_\mathrm{BLR}$. 
Given that the BLR is less luminous, the convection losses are dominant for electrons with $E' < 10$~MeV. Synchrotron cooling dominates during most of the JCI, except when the cloud has accelerated to $\Gamma_\mathrm{j}$. The jet radius is smaller, and as we are fixing the cloud initial size ($R_\mathrm{c,0}$), the expanded cloud maximum radius is almost half of the jet radius. This leads to a much larger fraction of the jet power being transferred to the shock. The average SED above $\epsilon > 1$~MeV is dominated by IC emission (below that synchrotron dominates). 

For a more powerful jet with $L_\mathrm{j} \sim 10^{48}$~erg~s$^{-1}$, we get $D \sim 10$ and the dominance of the BLR photon field in the radiative features is even more pronounced. However, in this case the cloud maximum radius is only $\sim 1\%$ of the jet radius, so the amount of jet power deposited in the shock is relatively small, and therefore the luminosity of a single JCI does not increase significantly with respect to the canonical value (Fig.~\ref{fig:L_vs_Lj}).

Assuming a constant $C_\mathrm{BLR}$, $N_\mathrm{c,tot} \propto f^{-2} \propto R_\mathrm{BLR}^2 \propto L_\mathrm{j}$. For a fixed value of $M_\mathrm{BH}$ \cite[supported by][]{Ghisellini2017}, $\bar{v}_\mathrm{c} \propto R_\mathrm{BLR}^{-1/2} \propto L_\mathrm{j}^{-1/4}$, and therefore the mean velocity of the clouds that are able to penetrate the jet is lower if $L_\mathrm{j}$ is high; for example, we obtain $\langle v_\mathrm{c,0} \rangle \approx 2 \times 10^9$~cm~s$^{-1}$ for $L_\mathrm{j} \sim 10^{45}$~erg~s$^{-1}$, and $\langle v_\mathrm{c,0} \rangle \approx 2 \times 10^8$~cm~s$^{-1}$ for $L_\mathrm{j} \sim 10^{48}$~erg~s$^{-1}$. Putting everything together, the scaling of $DC$ with $L_\mathrm{j}$ is weak and is about $DC \propto L_\mathrm{j}^{1/6}$. 

The average $\gamma$-ray luminosity and other properties of the JCI for different jet powers are summarised in Table~\ref{table:luminosities}. It is interesting to note that both the intensity of a single JCI and the duty-cycle of the events are only weakly dependent on $L_\mathrm{j}$. Therefore, the total (collective) luminosity of the JCIs is not very sensitive to $L_\mathrm{j}$, although more powerful jets are expected to be more luminous.

\begin{table}
 \caption{Comparison of the JCI properties for different jet powers. The mean expected luminosity in the 0.1--30~GeV energy range in the OF for a single JCI is $\langle \hat{L}_{0.1-30}\rangle$, whereas the total (collective) luminosity $\langle \hat{L}_\mathrm{tot}\rangle$ is estimated as $DC \times \langle \hat{L}_\gamma\rangle$.}
 \label{table:luminosities}
 \centering
 \resizebox{\columnwidth}{!}{%
 \begin{tabular}{cccccc}
 \toprule 
$L_\mathrm{j}$  & $z_\mathrm{j}$ & ${L'}_\mathrm{inj,peak}$ & $\langle \hat{L}_{0.1-30} \rangle$ & $DC$ & $\langle \hat{L}_\mathrm{tot} \rangle$ \\
 \, [erg~s$^{-1}$] & [cm] & [$L_\mathrm{j}$] & [erg~s$^{-1}$] & -- & [erg~s$^{-1}$] \\
 \midrule
$1 \times 10^{45}$  & $2\times 10^{16}$ & $5\times 10^{-3}$ & $2.8\times 10^{43}$ & $15$ & $4.2 \times 10^{44}$\\
$2.5\times 10^{46}$ & $1\times 10^{17}$ & $2\times 10^{-4}$ & $2.9\times 10^{43}$ & $34$ & $1.0 \times 10^{45}$\\
$1 \times 10^{48}$  & $7\times 10^{17}$ & $5\times 10^{-5}$ & $3.1\times 10^{43}$ & $53$ & $1.6  \times 10^{45}$\\
 \bottomrule
 \end{tabular}}
\end{table}   

Another aspect to take into consideration is the predicted X-ray luminosity, $L_\mathrm{X}$. As shown by \cite{Ghisellini2017} (their Fig.~6), FSRQs have a minimum in their SEDs close to 1~keV ($10^{17}$~Hz), and $L_\mathrm{X}$ ranges from $\sim 10^{43}$~erg~s$^{-1}$ to $\sim 10^{46}$~erg~s$^{-1}$ for $L_\mathrm{j}$ from $10^{45}$~erg~s$^{-1}$ to $10^{48}$~erg~s$^{-1}$, respectively. According to our model, for a typical JCI with $\eta_B = 1$ the expected $L_\mathrm{X}$ is $\sim 10^{43}$~erg~s$^{-1}$ regardless of $L_\mathrm{j}$, and the collective X-ray luminosity of multiple JCIs ranges from $\sim 10^{44}$~erg~s$^{-1}$ to $\sim 10^{45}$~erg~s$^{-1}$ for a value of  $L_\mathrm{j}$ from $10^{45}$~erg~s$^{-1}$ to $10^{48}$~erg~s$^{-1}$. Therefore, there is some tension between the predictions of our model when applied to FSRQs with weak jets. However, the expected value of $L_\mathrm{X}$ is sensitive to the adopted values of the model free parameters and it can be significantly lower than that obtained in the typical scenario (Sect.~\ref{sec:free_parameters}). A detailed analysis of the X-ray emission from JCIs is beyond the scope of this work.



\section{Conclusions}\label{sec:disc}


We have shown that under typical conditions JCIs conspicuously stand out as an efficient $\gamma$-ray emitting mechanism in blazars, although the evolution of an interaction between an AGN jet with a BLR cloud is complex. With a simple model accounting for cloud expansion and relativistic effects, we have shown that the interactions of BLR clouds with the jet in a blazar can produce significant $\gamma$-ray emission if $\xi_\mathrm{e} \gtrsim 0.1$ and if the BLR geometry is roughly spherical. This $\gamma$-ray emission is expected to be rather persistent, with some moderate fluctuations. Even though JCIs do not seem to be the dominant emitting mechanism, this mechanism cannot be neglected \textit{\emph{a priori}}  as partially shaping the observed SED of BLR-hosting blazars, in particular in the GeV range. 

If JCIs in the BLR of blazars do not significantly contribute to the emission seen by \textit{Fermi} from these sources \citep[as proposed by][due to the lack of strong attenuation in their spectra]{costamante18}, we can indirectly assess the study of the BLR properties (e.g. discarding a spherical shape) if we assume  that electron acceleration is efficient. Radio and optical variability can be used to test this \citep{Romero2000,Romero2002}. Otherwise, electron acceleration should be strongly suppressed (i.e. $\xi_\mathrm{e} \ll 1$). 

An improved model for BLR cloud-jet interactions (and for JCIs in general) should consider that the emission needs to be calculated in the shocked fluid frame rather than in the accelerating cloud reference frame. This improvement requires a more detailed analysis of the emitting fluid \citep[e.g.][]{delaCita2016}. Another improvement would require a more realistic BLR prescription, although this may not be an easy task given the uncertainties on BLR characterisation.


\begin{acknowledgements}

We thank the anonymous referee for the detailed revision of the manuscript and for the constructive suggestions. This work is supported by CONICET (PIP2014-0338) and ANPCyT (PICT-2017-2865), and also
by MDM-2014-0369 of ICCUB (Unidad de Excelencia `Mar\'{i}a de Maeztu'), and the Catalan DEC grant 2017 SGR 643. SdP acknowledges support from PIP 0102 (CONICET). V.B-R. and G.E.R acknowledge support from the Spanish Ministerio de Econom\'{i}a y Competitividad (MINECO/FEDER, UE) under grant AYA2016-76012-C3-1-P, with partial support from the European Regional Development Fund (ERDF/FEDER).

\end{acknowledgements}

\bibliographystyle{aa} 
\bibliography{biblio} 

\end{document}